\documentclass[aps,floats,floatfix,showpacs,amssymb,prd,twocolumn,superscriptaddress,nofootinbib,longbibliography,reprint]{revtex4-1}

\usepackage{amssymb,amsmath,verbatim,mathtools,needspace,enumitem,etoolbox,graphicx,physics,microtype,afterpage,xspace,tabularx,lmodern,multirow}
\usepackage{gensymb}
\usepackage[normalem]{ulem}
\usepackage[dvipsnames, usenames]{xcolor}
\definecolor{linkcolor}{rgb}{0.0,0.3,0.5}
\usepackage[unicode, colorlinks=true, linkcolor=linkcolor, citecolor=linkcolor, filecolor=linkcolor, urlcolor=linkcolor, linktocpage, breaklinks]{hyperref}
\usepackage[all]{hypcap}
\usepackage[T1]{fontenc}
\usepackage[utf8]{inputenc}
\usepackage[usenames,dvipsnames]{xcolor}
\hypersetup{colorlinks=true,citecolor=romared,linkcolor=romared,urlcolor=romared}
\definecolor{romared}{RGB}{142,0,28}
\usepackage{aas_macros}
\usepackage{makecell}
\usepackage{soul}
\usepackage{booktabs}

\usepackage{fontawesome}
\newcommand{\linkmaster}{\href{https://zenodo.org/records/17918405}{\scriptsize \faDownload}}

\usepackage{titlesec} 

\usepackage{titlesec}

\titleclass{\subsubsubsection}{straight}[\subsubsection]
\newcounter{subsubsubsection}[subsubsection]
\renewcommand\thesubsubsubsection{\thesubsubsection.\arabic{subsubsubsection}}
\titleformat{\subsubsubsection}
  {\normalfont\normalsize\bfseries}{\thesubsubsubsection}{1em}{}
\titlespacing*{\subsubsubsection}{0pt}{3.25ex plus 1ex minus .2ex}{1.5ex plus .2ex}
\newcommand{\OU}{\affiliation{Department of Physics, Oakland University, Rochester, Michigan 48309, USA}}
\begin{document}

\title{The Merger Rate of Primordial Black Holes} 

\begin{abstract}
The merger rate of primordial black hole (PBH) binaries can be used to understand the
source population of the merging black hole binaries observable through gravitational waves
(GWs) and also to constrain the possible contribution of PBHs to dark matter. In the
literature, the PBH merger rate is calculated analytically, assuming that PBH binaries stay in
isolation (i.e., are unperturbed) and evolve solely via GW emission during their entire lifetime.
However, if some or all of dark matter consists of PBHs, then as cosmic structures grow, PBH
binaries and single PBHs fall inside dark matter halos. In those halos, the PBH binaries'
interactions with their environment significantly affect the subsequent evolution of their orbital
properties. In this paper, we  present a numerical framework that accurately calculates the total
PBH merger rate by combining the evolution of isolated binaries outside halos with the dynamics
of binaries inside halos. In our work, we have found that the isolated binary channel is suppressed
at low redshifts, and dynamical interactions  in halos reshape the merger rate evolution with time, accelerating some mergers. 
At redshifts of $\lesssim 2$ the total merger rate is a factor of $\simeq 50 \%$ higher than the results assuming that all PBH binaries effectively stay unperturbed until their merger.  
Our simulations provide a definitive calculation on the total PBH merger rates, which are currently being probed and constrained from gravitational-wave
observations. We make our merger rates publicly available at Zenodo~\linkmaster
\end{abstract}

\author{Muhsin Aljaf}
\email{muhsinaljaf@oakland.edu}
\OU
\author{Ilias Cholis}
\email{cholis@oakland.edu}
\OU
\date{\today}
\maketitle

\section{Introduction}
Since the first detection of gravitational waves (GWs) from binary black hole mergers, by the LIGO-Virgo-KAGRA (LVK) collaboration~\cite{LIGOScientific:2016aoc, KAGRA:2021duu, KAGRA:2021vkt}, primordial black holes (PBHs) in the stellar mass range have regained attention \cite{Sasaki:2016jop, Bird:2016dcv, Carr:2016drx}. PBHs may have formed before matter-radiation equality in the early universe, making them a candidate for dark matter~\cite{Hawking:1971ei,Zeldovich:1967lct}. While most of the LVK-detected black hole merger events are of conventional astrophysical origin (see e.g.~\cite{KAGRA:2021duu, Kritos:2020fjw, Kritos:2022ggc, Bouhaddouti:2024ena}), the presence of PBH mergers in the observations can not be excluded. If even a fraction of these events is primordial, then LVK data can be used to constrain the PBHs abundance and contribution to dark matter~\cite{Bird:2016dcv, Sasaki:2016jop, Kovetz:2016kpi, Kovetz:2017rvv, Takhistov:2017bpt, Ali-Haimoud:2017rtz, Bellomo:2017zsr, Berti:2019xgr, Gow:2019pok, Hall:2020daa, Andres-Carcasona:2024wqk, Bouhaddouti:2025ltb}.

PBHs can exist in bound binaries and based on when these binaries formed, they are classified into two categories: (i) early binaries that form shortly after PBH production during the radiation-dominated era via gravitational decoupling of PBH pairs from the Hubble flow and (ii) late binaries that form through GW capture during close encounters of PBHs inside dark matter halos at later times (for a review on all formation channels, see Ref.~\cite{Raidal:2024bmm}).
The present-day merger rate of early binaries can be calculated analytically~\cite{Sasaki:2016jop,Ali-Haimoud:2017rtz}, assuming these binaries evolve in isolation through GW emission across their lifetime, having small orbital eccentricity, and neglecting environmental effects such as interactions with other PBHs inside dark matter halos. These assumptions may be valid for the approximate evaluation of their merger rates at high redshifts when PBH binaries are indeed in isolation. However, at later times, these binaries will fall into dark matter halos where they will interact with other objects, including other PBHs, leading to either their hardening or breaking/ionization. In this work, we refer to the binaries that evolve in isolation as unperturbed binaries. 

In our past work~\cite{Aljaf:2024fru}, we focused on the evolution of PBH binaries inside halos starting at $z\simeq12$. We showed that dynamical binary-single interactions within dark matter halos significantly alter the merger rates of early binaries. However, in our approach, we did not consider their earlier evolution outside the dark matter halos. Furthermore, one needs to model the possible disruption of PBH binaries due to the passing of other PBHs. This work aims to bridge the gap between the approach of the isolated evolution of unperturbed PBH binaries in Ref.~\cite{Ali-Haimoud:2017rtz}, which neglects environmental effects altogether, and the work of Ref.~\cite{Aljaf:2024fru} that focused on the PBH binaries inside halos. The early PBH binaries naturally split into two evolutionary paths: binaries that remain outside halos evolving only through purely GW emission (unperturbed binaries), and binaries that form halos and may have interactions with third objects. 

There is a third contribution to the merger rate of PBH binaries: binaries that form through GW two-body captures inside halos during close encounters of single PBHs. That contribution can be calculated semi-analytically~\cite{Bird:2016dcv, Aljaf:2024fru}. These binaries merge rapidly due to their high eccentricities~\cite{Cholis:2016kqi}. Their rates are small compared to the binary-single channel, but do give the absolute minimum merger rate, as that rate is independent of the initial conditions of the PBH binaries' orbital properties or their abundance.

In this work, we present a numerically unified framework to derive the proper merger rates of PBHs by accounting for the fraction of PBH binaries outside halos that evolve in isolation and the PBH binaries that evolve inside halos. Our framework accounts for three distinct contributions to the total PBH merger rate: (i) unperturbed binaries that remain outside dark matter halos and evolve purely through GW emission, (ii) binaries undergoing binary-single interactions inside dark matter halos that experience hardening/softening/ionization effects, and (iii) late binaries formed through GW capture during close encounters in halos.

Our goal is to estimate the total comoving rate of PBH as a function of redshift by properly combining these three contributions, i.e,
\begin{equation}
R_{\textrm{total}}(z) = R_{\textrm{unperturbed}}(z) + R_{\textrm{binary-single}} + R_{\textrm{captures}}(z).
\end{equation}
Throughout this paper, we assume that PBHs constitute all of the dark matter (i.e., $f_{\textrm{PBH}} = 1$) unless stated otherwise. Limits from GW searches do give $f_{\textrm{PBH}} \ll 1$, but as the merger rates are proportional to $f_{\textrm{PBH}}^2$, the scaling is easy to re-evaluate as we show at the end of this work. We use the cosmological parameters from \textit{Planck}~\cite{Planck:2018vyg}.

Our paper is organized as follows: Section~\ref{PBH merger channels} details the three PBH merger channels. 
We present the methodology used to compute the merger rates of unperturbed binaries and of binaries that later reside inside dark matter halos, building on our previous work in Ref.~\cite{Aljaf:2024fru}. We also adopt the direct-capture merger rates derived in Ref.~\cite{Aljaf:2024fru}.
In our simulations, we track in total billions of PBH binaries to take into account for both the hardening and the softening effects binary-single interactions can have on the PBH binaries. 
The impact of such interactions is environment-dependent. 
They depend on the total mass of the dark matter host halo, the location of PBH binaries within that halo, the moment those PBH binaries became part of a dark matter halo and how that halo grew in time. Our simulations take all those effects into account, allowing us to provide a definitive calculation of the PBH merger rates.
In section~\ref{RESULTS}, we present our results for the merger rates. We find that at high redshifts the merger rate is dominated by the unperturbed binaries, as should be expected, with a gradual increase in the contribution to the total rate from binaries that further hardened inside dark matter halos. At redshifts $\lesssim 2$ the total merger rate is about a factor of $50\%$ higher than the usually quoted rates based on the approximation of all PBH merging binaries evolving unperturbed. In Section~\ref{Conclusions}, we give our conclusions and discuss future prospects. Our rates are publicly available at  \href{https://zenodo.org/records/17918405}{Zenodo}.

\section{PBH merger  channels}\label{PBH merger channels}
In this section, we compute comoving PBH merger rates from three channels: unperturbed binaries evolving in isolation, binaries experiencing binary-single interactions in halos, and GW captures from close encounters in halos.

\subsection{Unperturbed binaries}
Unperturbed binaries are PBH binaries that evolve in complete isolation via GW emission throughout their entire lifetime. These binaries never experience environmental effects such as interactions with other PBHs. The evolution of their semi-major axis $a$ and eccentricity $e$ follows the Peters-Mathews equations~\cite{1964PhRv..136.1224P},
\begin{eqnarray}
\label{eq:Evol_a_gw}
\frac{da}{dt}\bigg|_{\rm GW}&=&
-\frac{64}{5} \frac{G^3}{c^5 a^3} \left(m_1+m_2\right)\left(m_1 \cdot m_2\right) F(e) \, ,\\
\label{eq:Evol_e_gw}
\frac{de}{dt}\bigg|_{\rm GW}&=&-\frac{304}{15} \frac{G^3}{c^5 a^4} \left(m_1+m_2\right)\left(m_1 \cdot m_2\right) D(e),
\end{eqnarray}
where $G$ is the gravitational constant, $c$ the speed of light, and $m_1$ and $m_2$ are the component masses of the binaries. Functions $F(e)$ and $D(e)$ account for the eccentricity dependence of the orbital properties’ evolution (see appendix~\ref{Function}).

To compute the comoving merger rate, we sample $N_{\textrm{sample}} = 5\times 10^{5}$ PBH binaries at formation redshift $z=3400$, drawing initial parameters $(a, e)$ from the theoretically motivated joint distribution of Refs.~\cite{Franciolini:2021xbq,Kavanagh:2018ggo}(see also~\cite{Sasaki:2016jop,Ali-Haimoud:2017rtz}). We  evaluate the number of mergers at timesteps of every $dt = 200 , \textrm{Myr}$, while we numerically solve  Eqs.~\eqref{eq:Evol_a_gw}–\eqref{eq:Evol_e_gw} at every $dt_{\textrm{local}} = 0.5 , \textrm{Myr}$ (for $z>35$ we use smaller $dt_{\textrm{local}}$). A binary in our samples is considered merged when its semi-major axis satisfies $a \leq 3 R_{\textrm{s}}$, where $R_{\textrm{s}}$ is the Schwarzschild radius of the binary.

The comoving merger rate of unperturbed PBH binaries with a monochromatic mass distribution at $m_{\textrm{PBH}}$ during the time interval $(t, t+dt)$ is ,
\begin{eqnarray}
\label{ISOLATED_BINARIES}
R_{\rm unperturbed}(t) &=& f_{\rm DM \,outside}(t) \cdot f_{\rm PBH\,binaries}(t) \nonumber \\
&& \times \frac{f_{\rm PBH}^{2} \cdot \rho_{\rm DM} \cdot S(t)}{2 \, m_{\rm PBH} \, dt} \,,
\end{eqnarray}
with the fraction of binaries merging during $(t, t+dt)$ evaluated as $S(t) = N_{\textrm{merge}}(t)/N_{\textrm{sample}}$, where $N_{\textrm{merge}}(t)$ is the number of mergers occurring between $(t,t+dt)$ in our fixed sample.  $f_{\rm DM,outside}(t)$ quantifies the fraction of dark matter outside halos (see Appendix~\ref{f_outside} for details), and $f_{\textrm{PBH,binaries}}$ is the initial fraction of PBHs in binaries at formation. $\rho_{\textrm{DM}}$ represents the comoving dark matter density. Note that Ref.~\cite{Ali-Haimoud:2017rtz} calculates this rate analytically, making the assumption that the binaries are circular, or have very small eccentricities during their evolution. Instead, we use a complete numerical approach to solve the system of Eqs. \eqref{eq:Evol_a_gw}-\eqref{eq:Evol_e_gw}.

\begin{figure}[!htbp]
\includegraphics[width=0.99\linewidth]{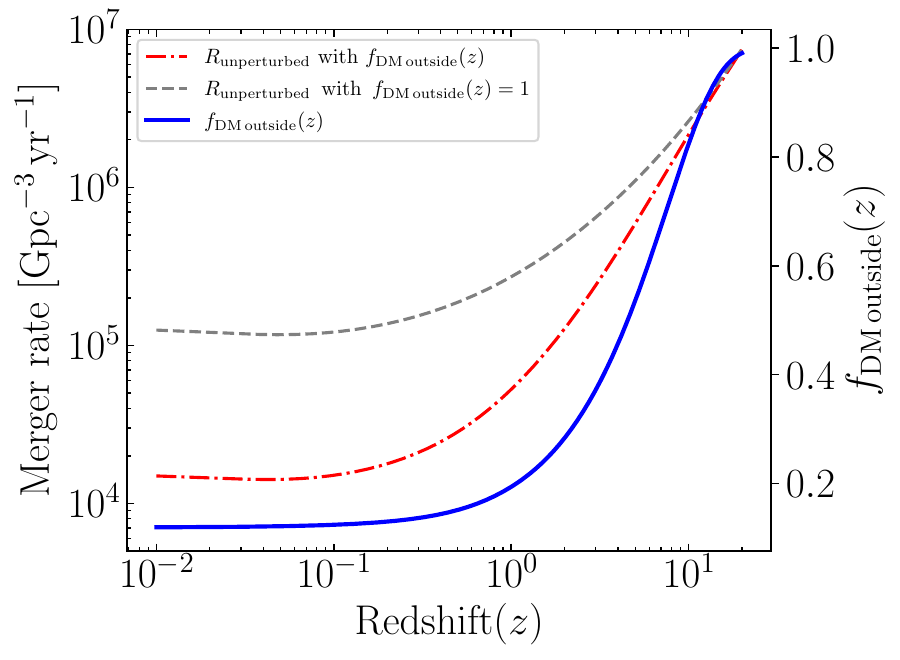}
\caption{The comoving merger rate of unperturbed early PBH binaries outside halos assuming $f_{\textrm{PBH binaries}}(z=3400)=0.5$ and $m_{\textrm{PBH}} = 30,M_\odot$. The red dashed-dotted curve shows the comoving merger rate of unperturbed binaries rescaled by the actual fraction of dark matter present outside halos $f_{\rm{DM,outside}}(z)$, while the gray dashed curve takes all PBH binaries to remain in isolation (unperturbed) setting $f_{\rm{DM,outside}}(z)=1$.  The blue solid  curve (right $y$-axis) shows the fraction of dark matter outside halos as a function of redshift.}
\label{fig:Iso_merger_rate}
\end{figure}

Using this setup, we evolve the PBH binaries from their formation taken to be at $z=3400$ to the present day $z=0$, covering an available time of $t_{\rm available} = 13.78~\textrm{Gyr}$. We present the results of our numerical simulation for the comoving merger rate of unperturbed binaries in Fig.~\ref{fig:Iso_merger_rate}, assuming an initial PBH binary fraction of $f_{\textrm{PBH,binaries}}(z=3400)=0.5$. The red dash-dotted curve shows the comoving merger rate of unperturbed binaries calculated using Eq.~\eqref{ISOLATED_BINARIES}.
For comparison, the gray dashed curve represents the case where all PBH binaries remain unperturbed throughout their lifetime, corresponding to $f_{\textrm{DM,outside}}(z)=1$ in Eq.~\eqref{ISOLATED_BINARIES}. This case is consistent with the results of Ref.~\cite{Ali-Haimoud:2017rtz}\footnote{Note that our rate is lower by a factor of 2 than Ref.~\cite{Ali-Haimoud:2017rtz} due to the fact that we set $f_{\rm{PBH ,binaries}}=0.5$ instead of 1.}. The blue solid curve (right $y$-axis) shows   the fraction of dark matter (and thus PBH binaries) residing outside virialized halos as a function of redshift, $f_{\rm{DM,outside}}(z)$.
The suppression of the comoving merger rate due to the decreasing fraction of PBH binaries outside halos becomes important for $z \lesssim 10$.

\subsection{Binary-single interactions in halos}
 The binary-single interaction channel accounts for PBH binaries that fall into dark matter halos and experience environmental effects such as hardening/softening/ionizations. This processes starts around $z\approx12$ when the  unperturbed PBH binaries that have not merged and the single PBHs start to cluster and form dark matter halos. Inside the halos, the PBH binaries can interact with single PBHs and wide PBH binaries via binary-single interactions \footnote{While the interaction between two PBH binaries is not a binary-single interaction, in almost all instances, at least one of the binaries is wide enough that the tighter of the two binaries effectively interacts with the PBH closer to its center of mass.}. Such interactions depend on the local density and velocity dispersion of the region where the binary resides within the halo, which determines whether a binary is hard or soft relative to its environment. 

\subsubsection{Binary classification and its evolution}
The nature of binary-single interactions depends critically on whether a binary is classified as hard or soft relative to its local environment in the halo. Following \citet{Heggie:1975tg}, we define the critical semi-major axis for hard binaries as,
\begin{equation}
a \leq a_{\rm h}(r,t) = \frac{G \, m_1  \, m_2}{2  \,m_3 \, v_{\rm disp}^{\rm env}(r, t)^2},
\label{a_hard}
\end{equation}
where $v_{\textrm{disp}}^{\textrm{env}}(r, t)$ is the dispersion velocity of the halo at distance $r$ from the center of the halo at time $t$ and $m_3$ is the mass of the interacting third object PBH. 
Since $a_h$ depends on $r$ and $t$, binaries can transition from hard to soft as the halo evolves. The hard binaries become harder (shrink) under binary-single interactions, while soft binaries may soften (widen) or be disrupted (ionized) according to Heggie-Hills law~\cite{Heggie:1975tg}. 

Within a given halo, we divide the evolution of PBH binaries into three regimes: hard,  soft, and intermediate. The third of these regimes corresponds to binaries near the hard-soft boundary.
\begin{enumerate}
    \item  In the hard regime where $a \le a_h$, we evolve binaries via both binary-single hardening and GW emission as \cite{1987Binney, Heggie:1975tg},
    \begin{eqnarray}
    \frac{da}{dt} &=&-\frac{G\, H \,\rho_{\rm env}(r, t)}{v_{\rm disp}^{\rm env}(r, t)}  a^2  + \frac{da}{dt}\bigg|_{\rm GW}, \label{eq:evol_a_hard} \\
    \frac{de}{dt} &=&+\frac{G\, H\,K(r,t)\,\rho_{\rm env}(r, t)}{v_{\rm disp}^{\rm env}(r, t)}  a + \frac{de}{dt}\bigg|_{\rm GW}. 
    \label{eq:evol_e_hard}
    \end{eqnarray}
    $H$ is the hardening rate fixed at $H=7.6 \cdot B$  with $B=\sqrt 3/2$ and $K$ is eccentricity growth rate determined by three-body scattering experiments~\cite{Quinlan:1996vp, Sesana:2006xw}. The factor $B$ accounts for the fact that both the single PBHs and the PBH binaries have the same velocity dispersion at a given radius $r$ within a halo.
    $\rho_{\rm env}(r, t)$ is the mass density of the local environment (including both single PBHs and PBH binaries) where the binary resides. We assume that this density follows the Navarro–Frenk–White profile~\cite{Navarro:1995iw}.
    
    \item  In the intermediate regime where $a_h <a \le 1.81 a_h$,  we evolve the binaries purely via GW emission using Eqs.~(\ref{eq:Evol_a_gw}) and (\ref{eq:Evol_e_gw}). We treat such binaries as neither soft nor hard. There is no clear boundary between hard and soft binaries.
    %1.81 a_h= $a>G m_{1}/(1.1 \, v_{\rm disp}^{\rm env}(r,t)^2)$,

    \item Finally in the soft regime where $a>1.81 a_h$, binaries might be softened by interacting with other PBHs and binaries. We evolve such binaries by~\cite{1987Binney},
        \begin{eqnarray}
        \frac{da}{dt} &=& +16 \sqrt{\frac{\pi}{3}} \, B \, \frac{G\,\rho_{\rm env}(r, t)}{v_{\rm disp}^{\rm env}(r, t)} \cdot (\ln{\Lambda}) \cdot a^2  \nonumber \\
        &&+ \frac{da}{dt}\bigg|_{\rm GW} \; \; \textrm{and}
        \label{eq:evol_a_soft} \\
        \frac{de}{dt} &=& -16 \sqrt{\frac{\pi}{3}} \, B \, \frac{G\, K\,\rho_{\rm env}(r, t)}{v_{\rm disp}^{\rm env}(r, t)} \cdot  (\ln{\Lambda}) \cdot a \nonumber \\
        &&+ \frac{de}{dt}\bigg|_{\rm GW}, 
        \label{eq:evol_e_soft}
        \end{eqnarray}
        where
        \begin{equation}
        \Lambda=1.1 \, 
        \frac{a \,v_{\textrm{disp}}^{\textrm{env}}(r, t)^2}{G m_{\rm 1}}.
        \label{eq:CoulombFact}
        \end{equation}
    Soft binaries can also get disrupted or ionized by interacting with other single PBHs and PBH binaries. We estimate the ionization and non-disruption probabilities of such binaries without dynamical evolution (see Appendix~\ref{IONIZATION} for details). 
\end{enumerate}

\subsubsection{Halo effects on a single PBH binary}

\begin{figure}
    \centering
    \includegraphics[width=0.99\linewidth]{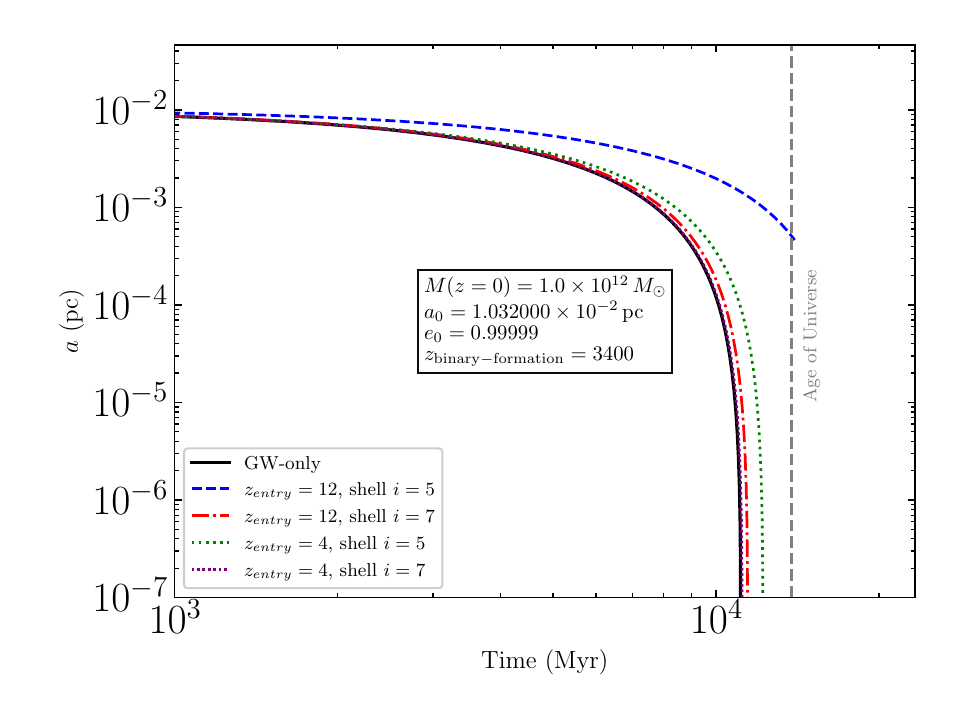}\\
    \vspace{-0.4cm}
    \includegraphics[width=0.99\linewidth]{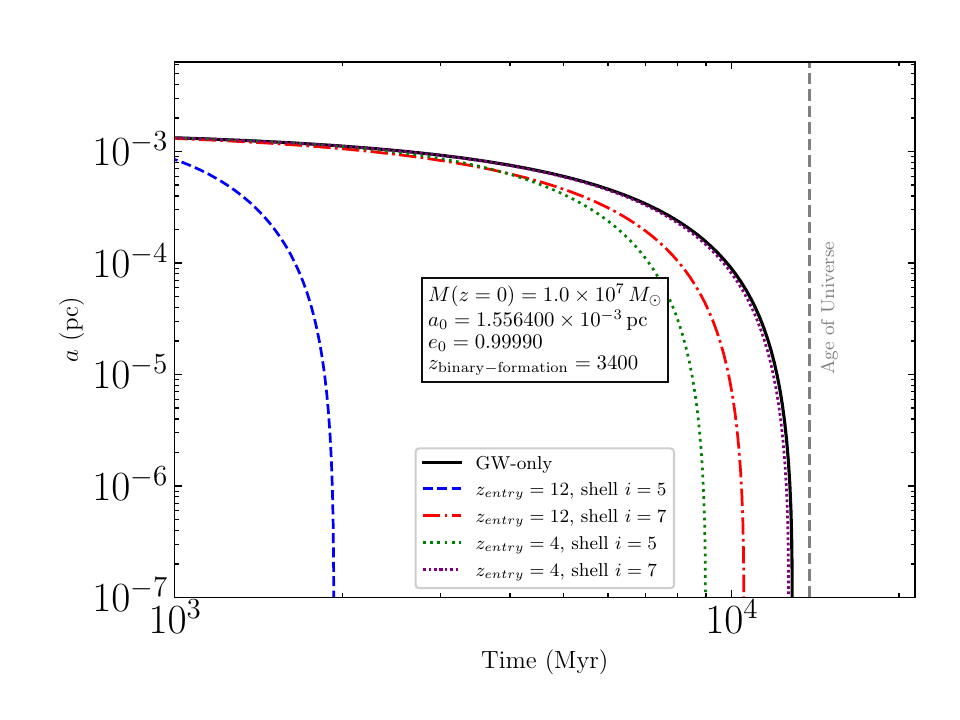}\\
    \vspace{-0.4cm}
    \includegraphics[width=0.99\linewidth]{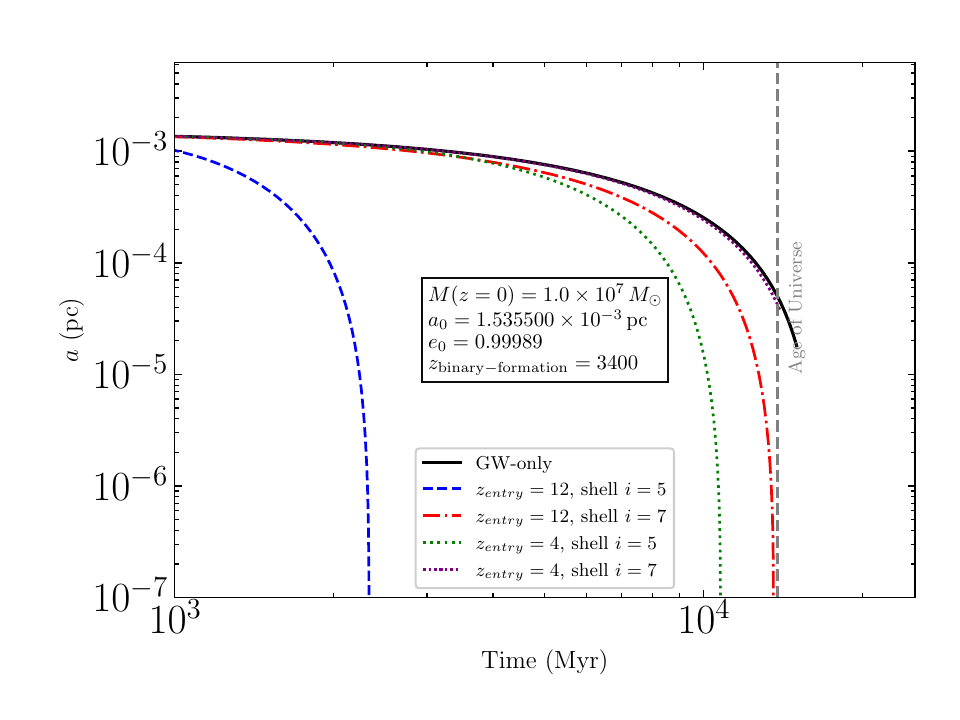}\\
    \vspace{-0.4cm}
    \caption{Examples (colored dashed, dotted, and dot-dashed lines) of how binary-single interactions inside dark matter halos can significantly alter the merger timescale for PBH binaries compared to the assumption that those binaries evolve in isolation (black solid lines). The top panel gives examples of binary-single interactions softening a binary and delaying its merger. The middle and bottom panels show instead examples of binary-single interactions causing hardening and accelerating a binary's merger. We give examples of different PBH binary's redshift of entry into the dark matter halo ($z_{\rm entry}$) and different local  environmental assumptions (shell of the dark matter halo). See text for more details.}
    \label{fig:softening&hardening}
\end{figure}

The evolution of a PBH binary is  determined by when it enters a halo and where within the halo it resides. In Fig.~\ref{fig:softening&hardening}, we present examples of a few  PBH binaries formed at $z=3400$, which initially evolved via GW emission (Eqs.~\eqref{eq:Evol_a_gw} and \eqref{eq:Evol_e_gw}) and then entered a halo at $z_{\rm entry}$. From $z_{\rm entry}$ to $z=0$, the binaries are placed in different halos based on the halo mass function at $z_{\rm entry}$. We show examples for binaries that entered a dark matter halo mass, which by redshift $z=0$ evolved into a $M=10^{12} M_\odot$ halo (top panel) and a $M=10^{7} M_\odot$ halo (middle and bottom panels). 
For such halos, we evaluate the environmental conditions  $v_{\rm disp}^{\rm env}(r, t)$ and $\rho_{\rm env}(r, t)$, in a sequence of ten spherical shells (see Ref.~\cite{Aljaf:2024fru} for more details). 

To see how the redshift of entry and local  environmental effects in the halo affect the binary evolution we show examples for binaries placed in some of the intermediate shells, $i=5$ and $i=7$ (at approximately 1/10th the Virial radius), entering those halos at early times ($z_{\rm entry} = 12$) in their growth, or at a time when the majority of dark matter is already inside halos ($z_{\rm entry} = 4$). 
Inside the halo, we evolve binaries  according to their type with respect to the halo: hard binaries via Eqs.~\eqref{eq:evol_a_hard}--\eqref{eq:evol_e_hard}, soft binaries via Eqs.~\eqref{eq:evol_a_soft}--\eqref{eq:evol_e_soft}, and intermediate binaries via GW emission alone.  
In the top panel, we show how for a PBH binary that would evolve only via GW emission and would merge at 11.1 Gyr (black line), softening can delay its merger to the point well beyond the age of the universe (blue dashed line). 
In the middle panel, we show how hardening can instead accelerate the merger timescale from 12.8 Gyr (black line) to a timescale that can be several Gyr less (as the blue dashed line). That has implications on the expected PBH merger rate at redshifts to be probed by future GW observatories.
Finally, in the bottom panel, we show how a PBH binary that via GW emission only would still not have merged (black line), may be merging just now (red dot-dashed line) or have already merged at a much earlier time (blue dashed line). 
Such binaries, in fact, are quite common and  responsible for the higher merger rate of PBHs that we find in this work compared to the calculation of purely unperturbed binaries (see \cite{Sasaki:2016jop, Ali-Haimoud:2017rtz}). 

\subsubsection{Halo Effects on Population of Merging PBH Binaries}

In this section, we study how halo environments affect the merger times of early PBH binaries. 
We simulate $5\times 10^{6}$ PBH binaries formed at $z=3400$, of which $9.1 \times 10^{4}$ will merge via GW emission in isolation, i.e., without any further binary-single interactions, between $z=12$ and $z=0$. 

\begin{figure}[!htbp]
    \centering
    \includegraphics[width=0.99\linewidth]{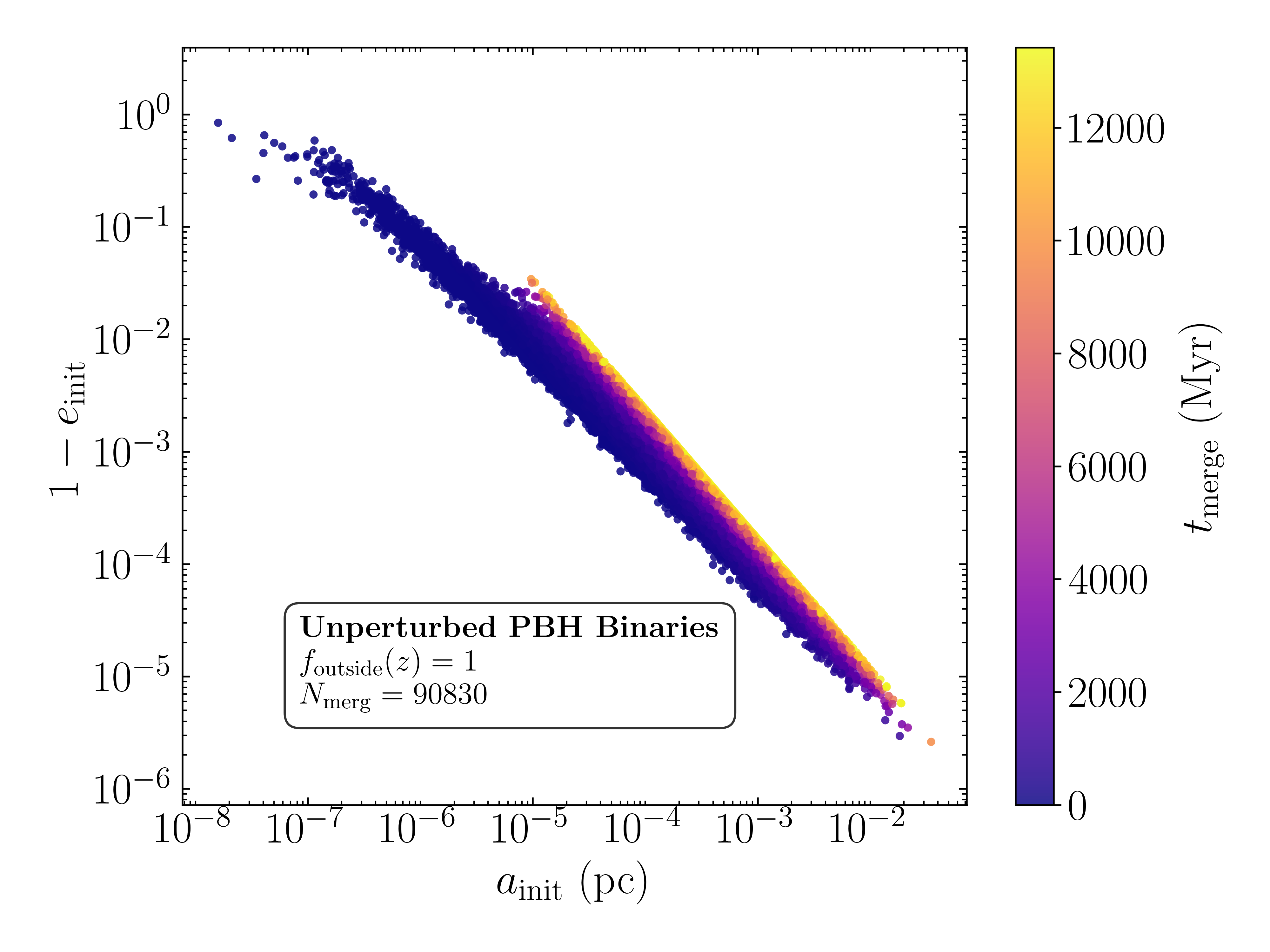}
    \vspace{-0.4cm}
    \includegraphics[width=0.99\linewidth]{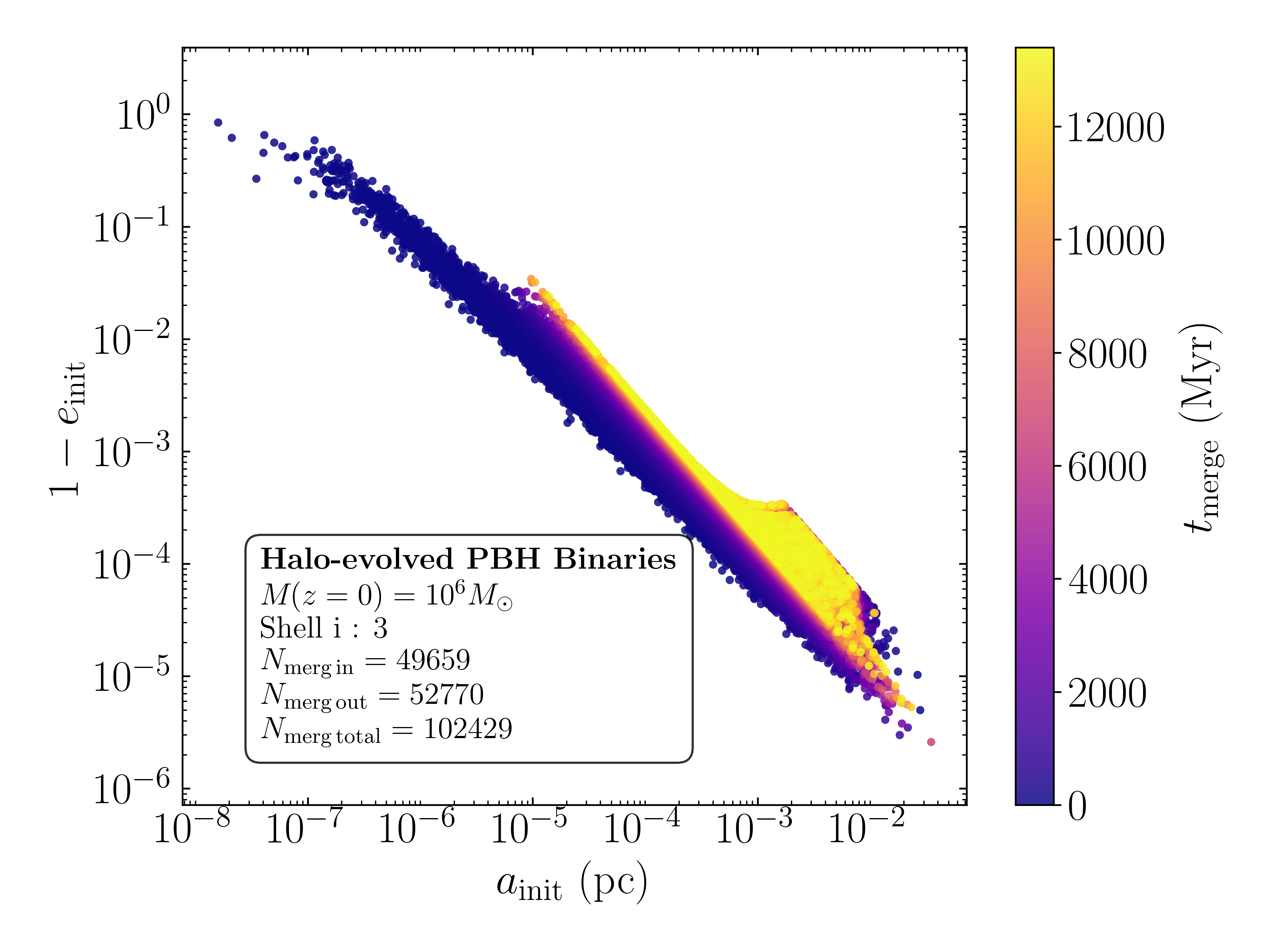}
    \vspace{-0.4cm}
    \caption{The distribution of the semi-major axes $a_{\rm init}$ and eccentricities $1-e_{\rm init}$ at $z=12$, for PBH binaries that merge by $z=0$. Top: all binaries are evolved in isolation via GW emission. Bottom: binaries evolved for simplicity all in a halo of mass $M=10^{6} M_\odot$ in one of its mid-radial distance shells ($i=3$ out of 5 shells). There are binaries that would merge over very long timescales ($t_{\rm merge} \sim 10^4~\rm Gyr$) in isolation but merge by $z=0$ due to dynamical interactions inside the halo. The binaries begin to fall inside the dark matter halo around $z= 12$ but continue to grow the halo up to $z=0$ (see text for more details).}
\label{fig:merger_scatter1}
\end{figure}

Fig.~\ref{fig:merger_scatter1}, shows the distribution of semi-major axes $a_{\rm init}$ and eccentricities $1-e_{\rm init}$ at $z=12$, for PBH binaries that formed at $z=3400$ and have merged by $z=0$. The subscript ``init'' refers to the initialization of the simulation at $z=12$, from which point on PBH binaries gradually form dark matter halos. 
The top panel displays binaries that evolve in isolation via GW emission and merge by today, i.e., assuming $f_{\rm  outside}(z) = 1$. The bottom panel shows binaries that merge by $z=0$, including the effects of binary-single interactions inside halos. 
For the bottom panel, we clarify again that our PBH binaries only gradually fall inside these dark matter halos  \footnote{Whether a binary enters a halo is determined probabilistically via $p_{\rm enter}$; for details, see  Eq.\eqref{P_enter} of Appendix~\ref{halo_effect}.}. 
Those binaries include binaries that would have merged anyway due to purely GW emission within the age of the universe, but also include binaries that 
in isolation would otherwise merge over a much longer timescales (in some cases even $\sim 10^4~\rm Gyr$).
Such binaries (yellow dots), mostly include binaries with high initial eccentricities, $1-e_{\rm init} < 10^{-3}$ and large initial semi-major axes, $a_{\rm init} > 10^{-4}$ pc. 
Binary–single interactions inside the halos affect the evolution of these binaries, increasing the fraction of binaries that merge by today.  
The color code in Fig.~\ref{fig:merger_scatter1}, gives the time of merger from the beginning of the simulation. Darker dots represent mergers occurring earlier in the simulation time. 

\begin{figure}[!htbp]
    \centering
    \includegraphics[width=0.99\linewidth]{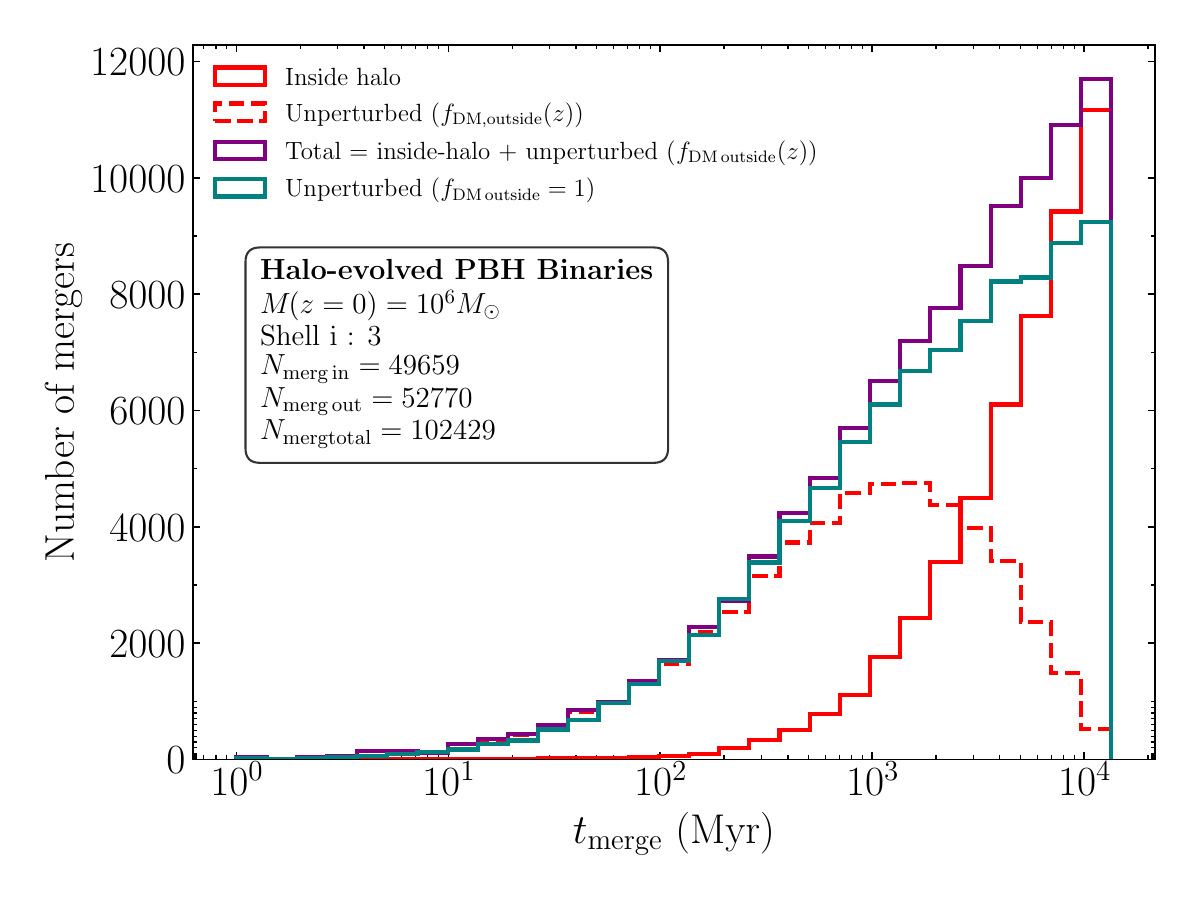}
    \vspace{-0.4cm}
    \includegraphics[width=0.99\linewidth]{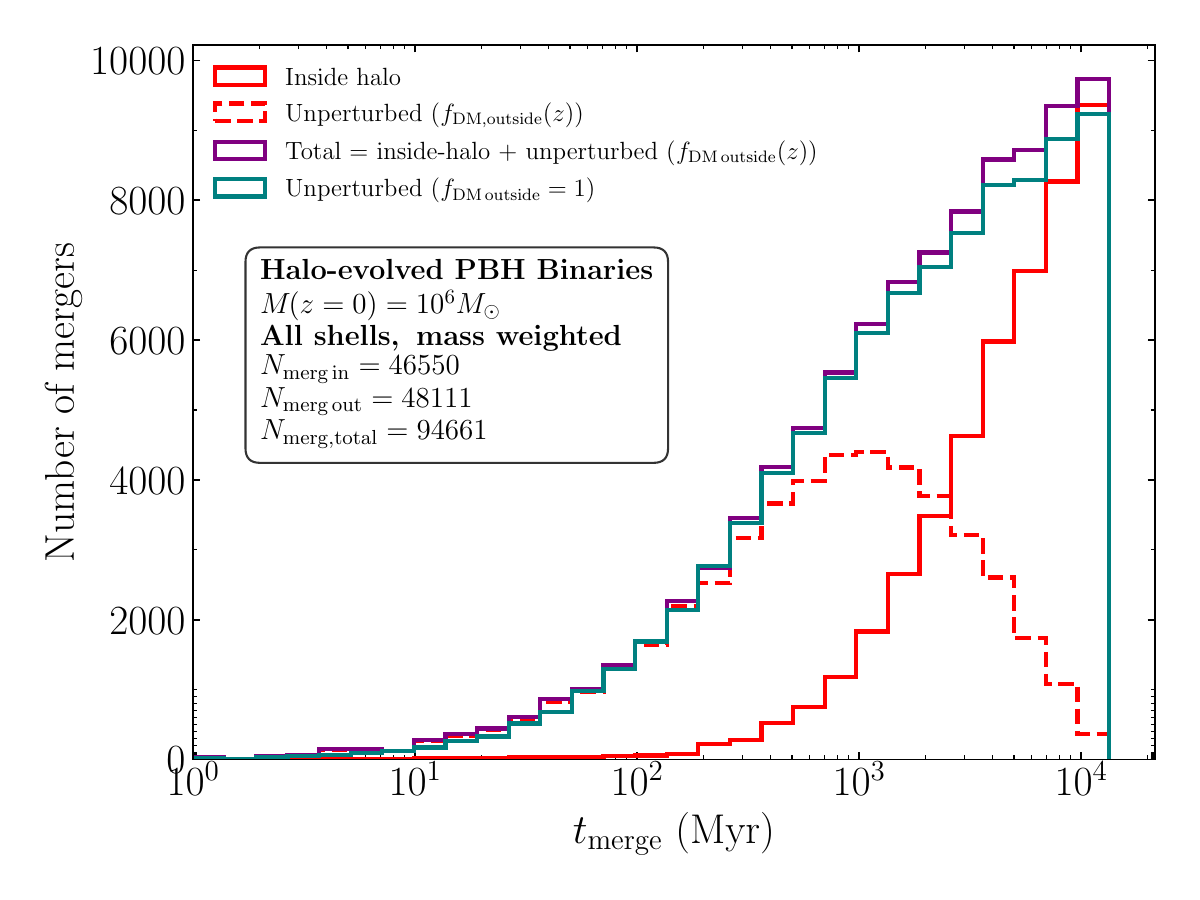}
    \vspace{-0.4cm}
       \caption{Histogram of the merger times for different PBH binary populations. We show results for $t_{\textrm{merge}}$ up to the age of the universe. Green: unperturbed binaries evolved in isolation, assuming $f_{\rm DM \, outside}(z)=1$. Solid red: binaries evolved inside halos, showing the acceleration of mergers due to binary–single  interactions. Dashed red: unperturbed binaries evolved in isolation properly weighted by $f_{\rm DM \, outside}(z)$. Purple: total population once the dark matter halo evolution is considered.}
\label{fig:merger_hist}
\end{figure}

Fig.~\ref{fig:merger_hist} presents histograms of merger times for PBH binaries, out of an initial sample of $5\times 10^{6}$ binaries formed at $z = 3400$ with the same original orbital distribution properties. Of those $4.75\times 10^{6}$ have not merged by $z=12$ when we start our simulations of the dark matter halo effects. 
The green histogram corresponds to unperturbed merged binaries assuming $f_{\rm  outside}(z)=1$, without any binary-single interactions inside any halo. There are 90830 such binaries (see also top panel of Fig.~\ref{fig:merger_scatter1}). 
For the remaining lines of these figures, for simplicity we take the PBH binaries to gradually fall inside a dark matter halo that by $z=0$ has grown to a mass of $M(z=0) = 10^{6} \, M_{\odot}$. 
In both panels, the red dashed line takes into account the evolving $f_{\rm DM \, outside}(z)<1$, and gives $N_{\rm merg \, out}=52770$ PBH binaries merging outside of the dark matter halo, i.e., before falling in. 
In the top panel, we show first the distribution of merger times assuming all PBH binaries experience the same environmental conditions once falling in, i.e., dark matter density and velocity dispersion of a mid-radial distance shell ($i=3$). For such an environment, there is a detectable (more than 10\%) increase in the total number of mergers most importantly at the last few Gyr of the simulation time.  
The red solid line gives the histogram of binary mergers taking place inside the halo. 
There are $N_{\rm merg \, in}=49659$ such binaries.
Finally, the purple line shows the combined histograms of PBH binary mergers outside and inside the dark matter halo $N_{\rm merg \, tot} = N_{\rm merg \, out}+N_{\rm merg \, in}=102429$. 
The hardening of binaries inside halos increases the total number of merged binaries and shifts the merger-time distribution to shorter times. At early times, the unperturbed simulation (green) and the full simulation (purple) agree, but starting at $z\approx 5$ the full simulation gives an increased number of PBH mergers. 
This shows the significance the local environments have in increasing the PBH merger rates.

In the bottom panel of Fig.~\ref{fig:merger_hist}, we account for the fact that our simulated dark matter halos are divided into several shells to account for the radially varying dark matter densities and velocity dispersions. This panel properly calculates the weight $w_{i}$ of the merger events of the binaries in each shell using $w_{i}=M_i/M_{h}$, where $M_i$ is the mass of shell $i$ and $M_{h}$ the total halo mass ($\sum_{i} M_i = M_{h}$). Even after accounting for the outer shells where the PBH binaries are mostly unperturbed, there is a detectable increase in the number of PBH mergers ($N_{\rm merg \, tot} = 94661$) due to binary-single interactions. In Appendix~\ref{halo_effect}, we give similar results to those of Figs.~\ref{fig:merger_scatter1} and~\ref{fig:merger_hist} for alternative halo environments. 

\subsubsection{Merger Rate Per Halo}
To estimate the comoving merger rate from binary-single interactions in halos of all masses, we first calculate the merger rate per halo. 
As we have mentioned earlier, the binaries' evolution in dense environments significantly depends on the local density and dispersion velocity using in Eqs.~(\ref{eq:evol_a_hard})-(\ref{eq:evol_e_soft}).
We follow the approach of our previous work~\cite{Aljaf:2024fru}, where we divide each halo into a number of $i$ mass shells  with characteristic radius $r=r_i$ from the center of the halo, evolving each shell's environmental properties with time according to the halo mass accretion history~\cite{Correa:2015kia} (see Ref.~\cite{Aljaf:2024fru} for details). %on shell division calculation and the estimation of local properties such as dispersion velocity and the density of each shell. 
Inside each shell of a given halo, we evolve a gradually increasing number of PBH binaries $N_{\textrm{sample},i}(t)$ from $z=12$ to today ($z=0$), over the available total time of $t_{\rm available} \simeq 13.43~\textrm{Gyr}$. This sample contains both hard, soft, and those binaries that are in the intermediate regime. 

Since the binary-single interactions can change the timescale to merger by several Gyr, either decreasing it or increasing it (see Fig.~\ref{fig:softening&hardening}), we evolve only the population of PBH binaries that would merge via GW emission only  (i.e., unperturbed) between $z=12$ and a forward evolution into the future of 100 times the age of the universe. Out of $5\times 10^{6}$ PBH binaries formed at $z=3400$, $2.49 \times 10^{5}$ binaries fall in that class. 
However, we keep track of the entire population of PBH binaries at all times. 
At any time, we record the number of binaries that are inside a halo versus outside. 
The probability for a PBH binary to be inside a spherical shell $i$ of dark matter mass $M_{i}$ is $f_{\textrm{inside}}(M_{i}, \, t)$. By tracking $f_{\textrm{inside}}(M_{i}, \, t)$, we can also track how many binaries enter a dark matter halo at each of its shells, at any given timestep. See Appendix~\ref{halo_effect}, for more details.
As the growth with time of dark matter halos depends on their final mass, these probabilities can not directly be evaluated simply by tracking $f_{\rm  outside} (z)$ shown in Fig.~\ref{fig:Iso_merger_rate}, which tracks the total amount of dark matter outside any halo of any mass (see for more details on that Appendix~\ref{f_outside}).

During the time interval $(t, t+dt)$, we count the number of binary mergers $N_{\textrm{merge},i}(t)$ and ionizations $N_{\textrm{ion},i}(t)$. The remaining binaries in the samples (that didn't merge or get ionized) are considered to ``survive'' $N_{\textrm{surv},i}(t)$ and will be further evolved. 
%Since our $N_{\textrm{sample},i}$  is larger than the actual population of the binaries exist shell, we need to rescale these sample results to the actual results that reflect the populations. 
The ratio of binaries that merged or got ionized during the step $dt$ at time $t$ in the shell $i$ is, 
%define a scaling factor for any event type $E$ (merger, ionization, survival) as the fraction of such events in the sample during the time interval
\begin{equation}
S_i^{(E)}(t) = \frac{N_{\textrm{event}=E,i}(t)}{N_{\textrm{sample},i}(t)},
\label{eq:SiE}
\end{equation}
where $N_{\textrm{event}=E,i}(t)$ is the number of events of type $E$ (merger or ionization) out of a sample with size $N_{\textrm{sample},i}(t)$ for shell $i$. Using these fractions, the actual merger/ionization rate of the simulation is,
\begin{equation}
R_i^{(E)}(t)=S_i^{(E)}(t) \times \frac{N_{\textrm{PBH binaries},i}(t)}{dt}.
\end{equation}
$N_{\textrm{PBH binaries},i}(t)$ is the actual number of PBH binaries inside shell $i$ at time $t$. This is evaluated by tracking the dark matter mass at each spherical shell $i$ in a given halo. The total merger rate per halo form binary-single interactions is obtained by summing the merger rates across all shells,
\begin{equation}\label{Rate_per_halo}
R_{\textrm{halo}}^{\textrm{binary-single}}(\rm t)=\sum_i R_i^{(\textrm{merge})}(\rm t).
\end{equation}

At each timestep, the actual number of binaries inside shell $i$, 
$N_{\textrm{PBH binaries},i}(t)$ is updated by accounting for the 
binaries that survived mergers or ionizations $S_i^{\textrm{surv}}(t) \times N_{\textrm{PBH binaries},i}(t)$
and adding new binaries entering from outside, $N_{\textrm{new},i}(t+dt)$ due to the mass growth of each shell. 
% The sample, $N_{\textrm{sample},i}(t)$, is also updated accordingly to reflect the relative proportions of survived and newly added binaries from outside the halo. For details on the update and sampling procedures, 
See Appendix~\ref{sample_update}. 

In Fig.~\ref{fig:B-S_rate_per_halo_single_halo_single halo}, we show the merger rate per halo as a function of redshift in units of $\textrm{yr}^{-1}$ for a halo with mass $\rm M(z=0)=1.15 \times 10^{12}\, M_\odot$. The dashed spiky blue curve is the merger rate per halo of binaries that undergo binary-single interactions, obtained from our numerical halo simulations. We use  Eq.\eqref{Rate_per_halo} to obtain the rate per halo. The solid line is the spline fit of the same curve.

\begin{figure}[!htbp]
    \centering
    \includegraphics[width=0.999\linewidth]{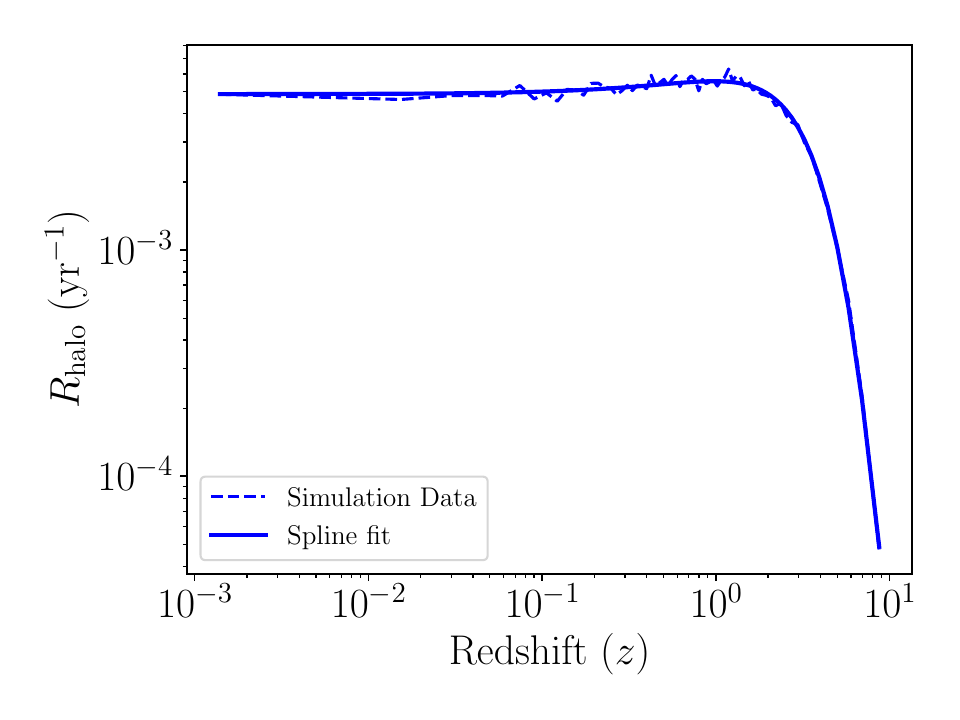}
    \caption{The merger rate per halo as a function of redshift for a halo with $\rm M(z=0)=1.15 \times 10^{12}  M_{\odot}$. We have assumed $f_{\textrm{PBH\, binaries}}=0.5$ and $f_{\mathrm{PBH}}=1$.}
    \label{fig:B-S_rate_per_halo_single_halo_single halo}
\end{figure}

We performed simulations for 45 halos with present-day masses between $10^4$ and $10^{15} \, M_\odot$ and obtained the merger rate of each halo. The comoving merger rate from binary-single interactions can be obtained via integrating over the  halo mass function~\cite{Press:1973iz},
\begin{equation}
\label{R_B_S}
R_{\rm{binary-single}}(z) =
\displaystyle \int_{M_{\rm min}}^{M_{\rm max}}
R_{\rm{halo}}^{\rm{binary-single}}(M, z) \,
\frac{dn(z)}{dM} \, dM \,,
\end{equation}
where $dn(z)/dM$ is the halo mass function giving the number density of halos per unit mass (see Eq.\eqref{eq:PS_hmf} in Appendix \ref{f_outside}). We take the limit of our integration from $M_{\rm min} = 10^{4} M_{\odot}$ and $M_{\rm max} = 10^{15} M_{\odot}$.
\subsection{GW Captures}
Two-body (direct) GW captures between single PBHs lead to the formation of new binaries (late binaries) with rapid subsequent mergers~\cite{Bird:2016dcv}. The comoving merger rate of  direct captures can be calculated semi-analytically and has a minor contribution to the total comoving merger rate of PBHs~\cite{Bird:2016dcv, Aljaf:2024fru}. 
For the direct capture rate, we use the updated result of~\cite{Aljaf:2024fru} that was roughly in agreement with Ref.~\cite{Bird:2016dcv}. 

\section{RESULTS}\label{RESULTS}
We provide the merger rate from each channel, namely: unperturbed binaries outside of dark matter halos, binaries that merge inside dark matter halos and undergo binary-single interactions, and binaries created from direct GW captures. We also show the proper total rate of PBHs.

\begin{figure}[!htbp]
    \centering
    \hspace{-0.0cm}
    \includegraphics[width=3.47in,angle=0]{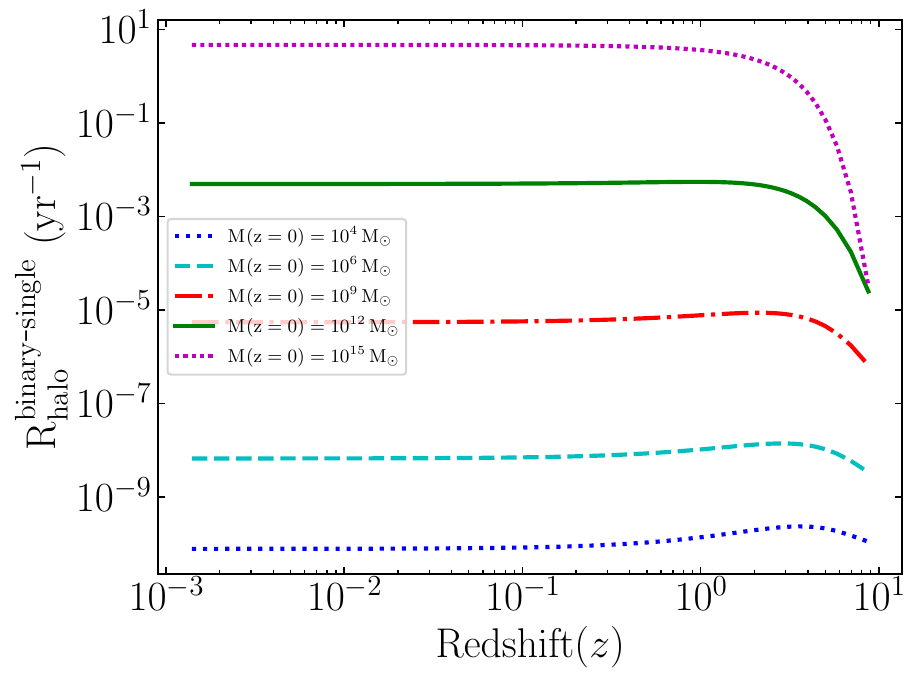}\\
    \includegraphics[width=3.48in,angle=0]{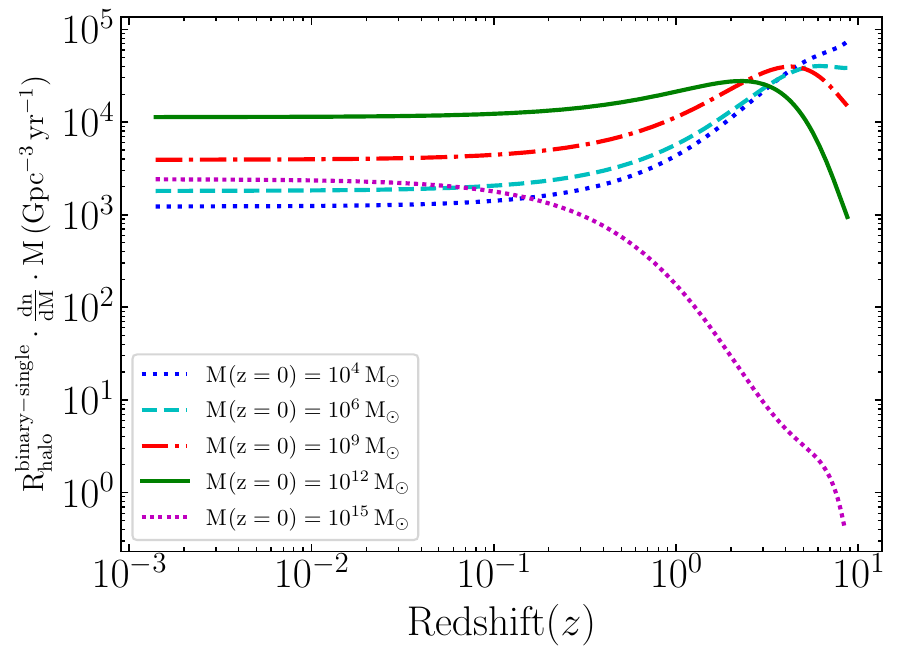}
    \vspace{-0.4cm}
    \caption{
    \textit{Top}: merger rate per halo via the binary–single interaction channel, for halos with present-day masses between $10^{4}$ and $10^{15}\,M_{\odot}$. Each of those halos grows in size with time (see text for details).
    \textit{Bottom}: contribution of each halo mass to the  comoving merger rate density, $R_{\rm halo}^{\rm binary-single}\, \times (M\cdot dn/dM)$, as a function of redshift. Low mass halos dominate at high redshift, while more massive halos increasingly contribute at later times, which reflects the hierarchical assembly of dark matter halos.}
    \label{fig:B-S_rate_per_halos}
\end{figure}

 In Fig.~\ref{fig:B-S_rate_per_halos} (top panel), we present the merger rate per halo of the binary–single-interaction channel for five different halos with present-day masses between $10^{4}$ and $10^{15} \,M_{\odot}$ assuming a monochromatic PBH mass distribution of $m_{\rm{PBH}} = 30 \,M_\odot$.
 We present how these rates evolve with redshift as the host halos grow.  
 As each halo's density and velocity dispersion profiles evolve in their own manner, the impact of the binary-single interaction terms in Eqs.~(\ref{eq:evol_a_hard})-(\ref{eq:evol_e_soft}) is affected differently. This is reflected to the per halo merger rates. For the smaller dark matter halos their per halo merger rate slightly drops from $z \simeq 3$ to today. 
 For a Milky Way sized dark matter halo (present mass of $10^{12} \, M_{\odot}$) the rate per halo stabilized after $z \simeq 2$, while in the most massive halos the rate continued to increase up to our era. 
 This causes deviations from a direct linear relationship in the merger rate per halo and its mass i.e. the number of contained PBH binaries.  

As the halo masses grow with time and dark matter halos merge with each other, the distribution of dark matter halo masses vastly evolves with time. In the bottom panel of Fig.~\ref{fig:B-S_rate_per_halos}, we represent the contribution of each of these halo masses to the comoving merger rate for the binary-single interaction channel.
That is we plot the $R_{\rm halo}^{\rm binary-single} \times  M\cdot dn/dM$, as a function of redshift, where $dn/dM$ is the number density of halos per mass per comoving volume.  Low-mass halos, which at early times contain most of the dark matter already inside halos, dominate the rate at high redshifts, while the more massive halos become increasingly important at later times, which shows the impact of hierarchical halo formation. In the current era, the rate is dominated by the contribution of halos with masses $M(z=0)$ of $10^{9} - 10^{13}M_\odot$, with however, a still important contribution from lower mass halos.

\begin{figure}[!htbp]
\centering
\includegraphics[width=0.99\linewidth]{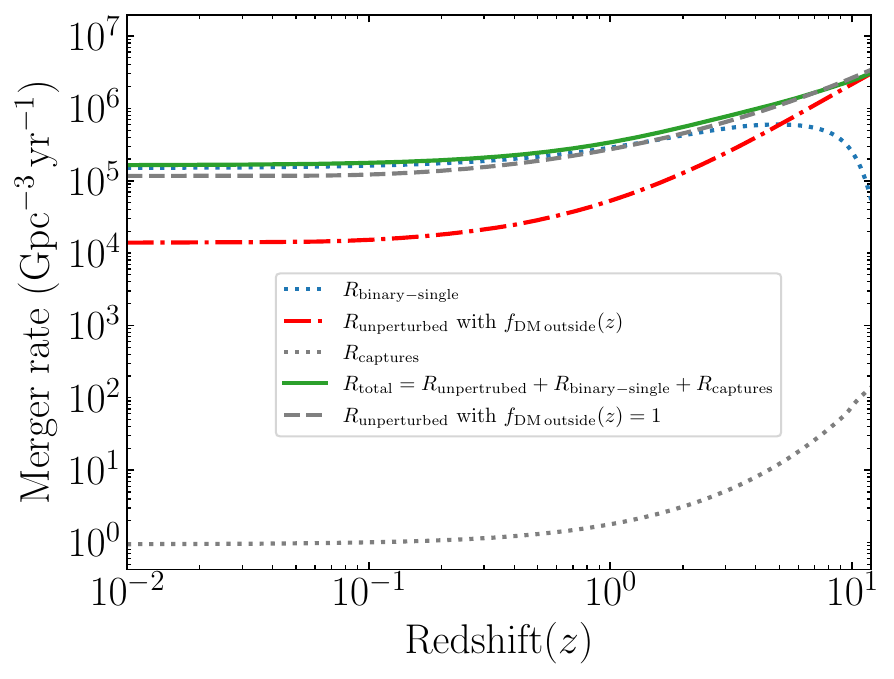}
\vspace{-0.4cm}
\caption{The comoving merger rate of PBH binaries as a function of redshift for the different channels. 
The red dash-dotted curve shows the rate from unperturbed binaries outside halos, accounting for the actual fraction of PBHs outside halos. The gray dashed curve, presented for comparison only, shows the unperturbed rate assuming all PBHs remain outside halos (consistent with Ref.~\cite{Ali-Haimoud:2017rtz}). The blue dotted curve shows the merger rate from binaries that also underwent binary–single interactions in halos. The gray dotted curve corresponds to mergers of late-time binaries formed via GW capture. The green solid curve represents the total comoving merger rate.}
\label{fig:Total_rates:channels}
\end{figure}

In Fig.~\ref{fig:Total_rates:channels}, we present the comoving merger rates from all three merger channels as a function of redshift.  
The red dash-dotted curve represents the proper comoving merger rate of unperturbed binaries that reside outside halos and evolve through GW emission only, over their entire lifetime. 
The rate is obtained assuming an initial binary fraction of $f_{\rm{PBH\,binaries}}(z=12)=0.5$. 
In that curve, the actual fraction of PBHs inside and outside of halos, $f_{\rm{DM\,outside}}(z)$ is properly accounted for.
For comparison only, the gray dashed curve shows the comoving merger rate of unperturbed binaries following the assumptions of Ref.~\cite{Ali-Haimoud:2017rtz}, which is equivalent to $f_{\rm{DM,outside}}(z)=1$ at all times. 
The red dash-dotted and the gray dashed curves are also shown in Fig.~\ref{fig:Iso_merger_rate}.
The blue dotted curve shows the comoving merger rate from binary–single interactions inside halos, including all halo masses from $10^{4} \, M_{\odot}$ to $10^{15} \, M_{\odot}$, obtained using Eq.~\eqref{R_B_S}. 
The gray dotted curve corresponds to the rate of late-time binaries that form via GW capture of single PBHs inside halos. This rate is calculated semi-analytically and adapted from~\cite{Aljaf:2024fru}. It gives only a tiny contribution to the total merger rate. 
Finally, the solid green curve shows the total comoving merger rate of PBHs, obtained by summing up all three channels using the unperturbed rate with the evolving $f_{\rm{DM\,outside}}(z)$. 
The comoving rates presented above are all obtained under the assumption that $f_{\rm{PBH}}=1$ and $f_{\textrm{PBH binaries}}=0.5$, which is very optimistic as the recent constraint from LVK data suggests an upper bound  $f_{\rm PBH} \sim 10^{-2}$ (see e.g. \cite{Bouhaddouti:2025ltb}). 
The total comoving merger rate $R_{\rm total}$ depends to $f_{\rm PBH}$ and
$f_{\textrm{PBH binaries}}$ as, 
\begin{eqnarray}\label{Rate_fpbh_rescaling}
R_{\rm total} &=& f_{\rm PBH}^2 [ R_{\rm captures} + 2 \, f_{\rm PBH\,\rm binaries} \nonumber \\
&& \times ( R_{\rm unperturbed} + R_{\rm binary-single} ) ].
\end{eqnarray}
The factor of 2 in that equation is because the $R_{\textrm{halo}}^{\textrm{binary-single}}$ and  $R_{\rm unperturbed}$ components are evaluated for $f_{\rm{PBH} \textrm{ binaries }}=0.5$. 
We note that the total PBH comoving merger rate scales as $f_{\textrm{PBH}}^{2}$ because forming binaries either at  early or late times requires pairs of PBHs.  
Moreover, only early merger channels (unperturbed and binary-single) depend on the binary fraction, $f_{\rm{PBH} \textrm{ binaries}}$ as their merger rate is proportional to the number of the early PBH binaries. 

\begin{figure}[!htbp]
    \centering
    \includegraphics[width=0.99\linewidth]{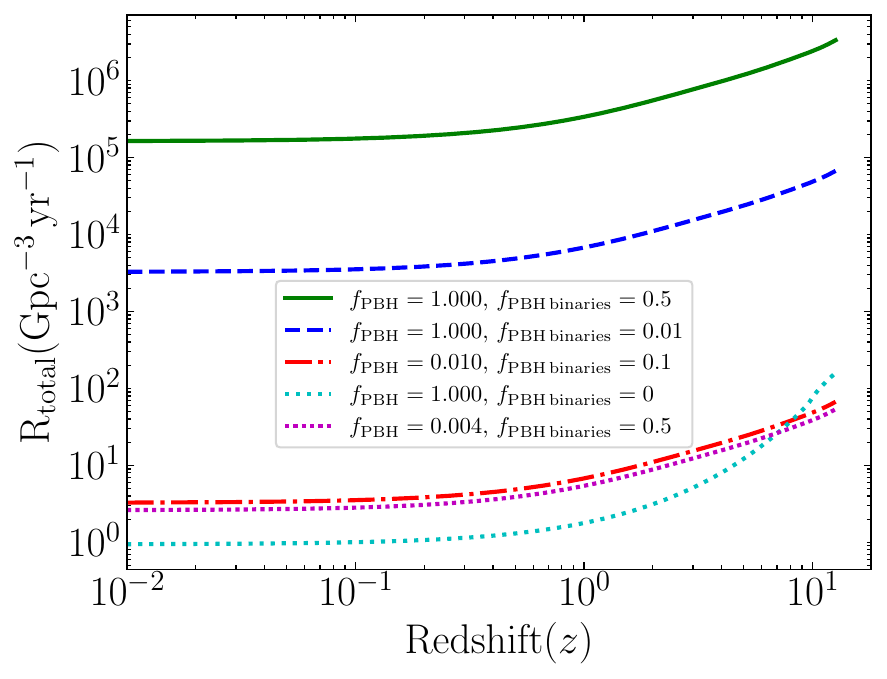}
    \caption{The total comoving merger rate $R_{\rm total}$ as a function of redshift $z$ for different PBH fractions. The five curves correspond to different choices of the  PBH binary fraction $f_{\rm PBH\,binaries}$ and the total PBH fraction $f_{\rm PBH}$. The lower three rates at $z<1$ are consistent with current LVK limits \cite{Bouhaddouti:2025ltb}.}
    \label{fig:fractions}
\end{figure}

In Fig.~\ref{fig:fractions}, we show the total comoving PBH merger rate $R_{\textrm{total}}$ as a function of redshift $z$, for five different  combinations of the PBH fraction $f_{\textrm{PBH}}$ and the PBH binary fraction $f_{\textrm{PBH binaries}}$, assuming that $f_{\textrm{PBH binaries}} + f_{\textrm{single PBH}} = 1$, where the latter fraction is the fraction of single PBHs. 

\begin{figure}[!htbp]
    \centering
    \includegraphics[width=0.99\linewidth]{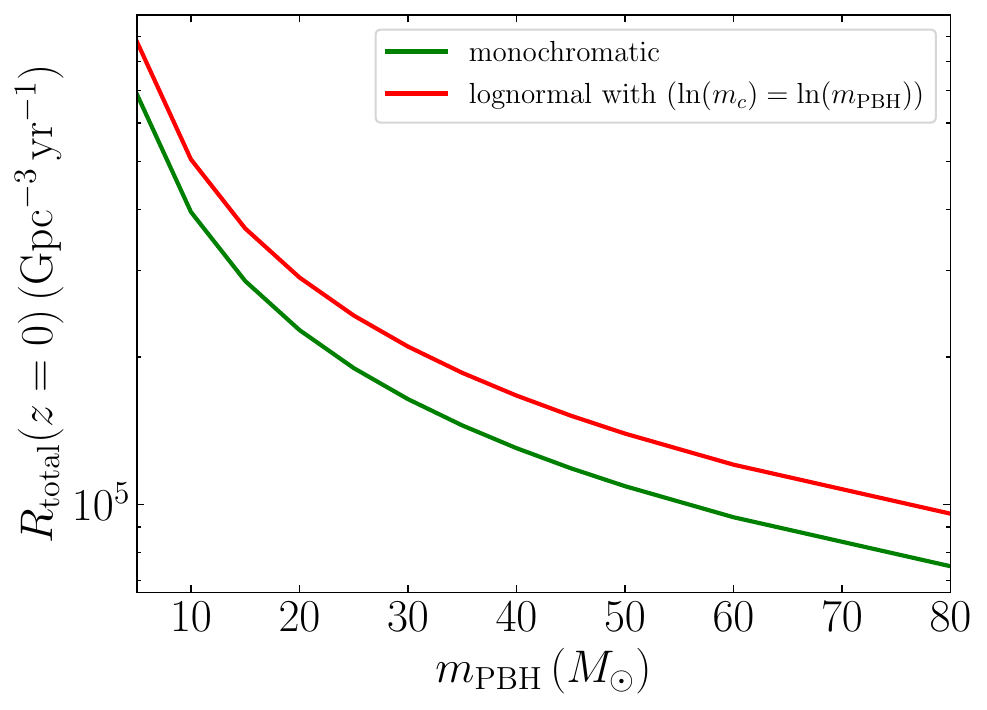}
    \caption{The total comoving merger rate $R_{\rm total}$  as a function of PBH mass $m_{\rm PBH}$ at $z=0$, assuming $f_{\rm PBH}=1$ and  $f_{\rm PBH\,binaries}=0.5$.}
    \label{fig:R_z0_mpbh}
\end{figure}

We find in our simulations that the total merger rate of a monochromatic distribution of PBHs with mass $m_{\rm PBH}$ scales as $R_{\rm total} (z) \propto m_{\rm PBH}^{-0.8}$. This is in good agreement with the previously derived parametric dependence from Refs.~\cite{Ali-Haimoud:2017rtz, Kavanagh:2018ggo}. We show this result in Fig.~\ref{fig:R_z0_mpbh}, for the merger rate at $z=0$ for $m_{\rm PBH}$ mass between 5 and 80 $M_{\odot}$.

\section{Conclusions}\label{Conclusions}

In this work, we present a numerical framework to calculate the total comoving merger rate of PBHs from their formation to the present day. We combine the merger rates from three distinct channels: (i) from unperturbed binaries evolving purely through GW emission, in isolation, outside dark matter halos, (ii) from binaries that fall inside halos of different masses and undergo dynamical binary-single interactions inside those halos, and (iii) from late-time binaries formed via GW capture interactions inside halos. 

We find that the merger rate of unperturbed binaries  starts getting suppressed around redshifts $z \lesssim 10$  as the fraction of dark matter outside halos starts to decrease and dark matter begins to cluster and form halos (shown in Fig.~\ref{fig:Iso_merger_rate}). By redshift of $z\simeq 1$, unperturbed binaries contribute only about $20 \%$ of the total merger rate, with that contribution falling to about $10 \%$ at by $z=0$. 

In fact, the main contribution to the total PBH merger rate comes from binaries inside halos starting from $z\simeq 5$ (see Fig.~\ref{fig:Total_rates:channels}).
While PBH binaries merging at $z\simeq 5$ have been mostly unperturbed up to the time of their merger, binaries that merge later, have enough time for their orbital properties' evolution to be affected by their respective environments inside the halos, described by Eqns.~(\ref{eq:evol_a_hard})-(\ref{eq:evol_e_soft}), as we show in Figs.~\ref{fig:softening&hardening}, \ref{fig:merger_scatter1} and~\ref{fig:merger_hist}. 
We simulate the gradual growth of dark matter halos following their evolution with redshift from Ref.~\cite{Correa:2015kia}, which affects the halo environments, i.e., the time-dependent values of dark matter density and velocity dispersion. 
Some PBH binaries' orbital evolution may be delayed as a result of PBH binary-single interactions, while more commonly these interactions accelerate the evolution of the binaries' orbital properties. Examples of individual binaries evolution are provided in Fig.~\ref{fig:softening&hardening}. 

We calculate the merger rate of PBHs inside dark matter halos for different halo masses between $10^{4} \, M_{\odot}$ and $10^{15} \, M_{\odot}$ and combine the contribution from all halos by properly accounting for the time-evolving halo mass function \cite{Press:1973iz, Murray:2013qza} (see Figs.~\ref{fig:B-S_rate_per_halo_single_halo_single halo} and~\ref{fig:B-S_rate_per_halos}). At redshifts $\sim 5-10$ the merger rate from inside halos is dominated by the smaller mass halos, as expected given that most dark matter is in such halos.
However, from redshift of $\simeq 2$, the halos that will grow by $z=0$ to a mass of $10^{9} - 10^{13} \, M_{\odot}$, provide the dominant host halo environments for the PBH mergers. 
Such an effect can inform future searches for possible cross-correlation signals between gravitational-wave detections and galaxy catalogues \cite{Raccanelli:2016cud}.  

Our work shows that including environmental effects to account for binary-single interactions is important for an accurate PBH merger rate calculation. 
Adding to the merger rate inside halos the component from late-formed direct captures and from mergers of binaries outside halos, we provide a definitive calculation of the total PBH merger rate up to a redshift of 10 (shown also in Fig.~\ref{fig:Total_rates:channels}).
At redshifts of $z \lesssim 2$, we find a total merger rate that is a factor of $50 \%$ higher compared to the assumption that merging PBH binaries evolve from their formation to their merger effectively unperturbed (e.g. Ref.~\cite{Ali-Haimoud:2017rtz}). 
Our calculated rates can be easily rescaled to account for different assumptions on the abundance of PBHs and the abundance of initially formed PBH binaries, as we show in Fig.~\ref{fig:fractions} and Eq.~(\ref{Rate_fpbh_rescaling}), and can be used to derive constraints on the abundance of PBHs from ongoing and future gravitational-wave observatories. 
We calculate the PBH merger rates for a monochromatic mass distribution with $m_{\rm PBH}$ between 5 and 80 $M_{\odot}$ and also for log-normal distributions peaking in the same mass range (see Fig.~\ref{fig:R_z0_mpbh}).
Our results on the PBH merger rates are made publicly available through \href{https://zenodo.org/records/17918405}{Zenodo}.

\acknowledgments
The authors would like to thank David Garfinkle for his valuable input during this project. MA and IC are supported by the National Science Foundation, under grant PHY-2207912.
We acknowledge the use of the publicly‑available Python package \texttt{HMFcalc} \cite{Murray:2013qza} for computing halo mass functions.

\appendix
\section{Functions for the evolution of the orbital properties of black hole binaries}\label{Function}
The functions $F(e)$ and $D(e)$ appearing in Eqs.~(\ref{eq:Evol_a_gw})--(\ref{eq:Evol_e_gw}) are defined as,
\cite{1964PhRv..136.1224P},
\begin{equation}
F(e)=\left(1-e^2\right)^{-7 / 2} \cdot\left(1+\frac{73}{24} e^2+\frac{37}{96} e^4\right) ,
\end{equation}
and
\begin{equation}
D(e)=\left(1-e^2\right)^{-5 / 2} \cdot\left(e+\frac{121}{304} e^3\right).
\end{equation}

\section{Fraction of unperturbed binaries outside dark matter halos}\label{f_outside}
Ref.~\cite{Ali-Haimoud:2017rtz}, assumed that all early binaries evolve in isolation across redshift, either because they are not part of dark matter halos or because most of the dark matter is at the outer parts of halos, where the chance of interaction with third objects is small. While on average at high redshifts this is valid, in our work, we model the evolution of all PBH binaries that are within the Virial radius of a dark matter halo and can also interact with third objects (other PBHs) in those environments. 
Thus, in order to avoid double counting, we need to estimate the fraction of dark matter that is outside halos, $f_{\rm{DM\,outside}}(z)$, i.e., at distances further away than the halo's Virial radius at any given redshift $z$. We define, 
\begin{equation}
f_{\rm{DM\,outside}}(z)=1-f_{\textrm {DM\,inside}}(z),
\end{equation}
where $f_{\textrm {DM\,inside}}(z)$ the fraction of dark matter inside dark matter halos,
\begin{equation}
f_{\textrm {DM\,inside}}(z) = \frac{1}{\rho_{\rm m}} \int_{M_{\rm min}}^{M_{\rm max}} M \frac{dn}{dM}(z) \, dM,
\end{equation}
where $\rho_{\textrm{m}}$ is the comoving matter density and $dn(z)/dM$ is the halo mass function giving the comoving number density of halos per unit mass.  We take $M_{\rm min} = 10^{4} M_{\odot}$ and $M_{\rm max} = 10^{15} M_{\odot}$ which corresponds to the typical range of halo masses that could be composed entirely of PBHs. We use the Press-Schechter halo mass function~\cite{Press:1973iz},
\begin{equation}
\label{eq:PS_hmf}
\frac{dn}{dM} = \frac{\rho_{\rm m}}{M} \left| \frac{d \ln x^{-1}}{dM} \right| f(x),
\end{equation}
with 
\begin{equation}
f(x) = \sqrt{\frac{2}{\pi}} \frac{\delta_c}{x} \exp\left(-\frac{\delta_c^2}{2 x^2}\right),
\end{equation}
where $x$ is the variance of the linear matter density field and $\delta_c = 1.686$ is the critical collapse threshold. The mass function is computed using the publicly available Python package \texttt{HMFcalc}~\cite{Murray:2013qza}.

The redshift evolution of $f_{\textrm{DM \,outside}}(z)$ is shown in Fig.~\ref{fig:Iso_merger_rate} as a solid blue curve (right $y$-axis).  We note that around $z\simeq 10$ the dark matter halo formation starts being important and the fraction $f_{\textrm{DM\,outside}}(z)$  starts to decrease rapidly.  

\section{Ionization of soft binaries}\label{IONIZATION}
Binaries with $a > 1.81 a_{h}$ are soft and can be disrupted by encounters with other PBHs through two processes, evaporation and ejection \cite{1987Binney}. Evaporation occurs when the binary gradually gains energy from multiple distant encounters until it disrupts. The timescale of this process is,
\begin{equation}
t_{\textrm{evap, $i$}}\left(a, r_i,t\right)= \frac{\sqrt{3} \, v_{\textrm{disp}}^{\textrm{env}}(r_i,t)}{16 \, \sqrt{\pi}\, B\,\, \rho_{\textrm{env}}(r_i, t) \, G \, a \cdot \ln \Lambda_{\textrm{bin, $i$}}}.
\end{equation}
Ejection happens when a single close encounter disrupts the binary. The timescale of such an encounter is,
\begin{equation}
t_{\textrm{ejc, $i$}}\left(a, r_i,t\right)= \frac{3\sqrt{3} \, v_{\textrm{disp}}^{\textrm{env}}(r_i,t)}{80 \, \sqrt{\pi}\, B\,\,  G \rho_{\textrm{env}}(r_i, t)  \, a }.
\end{equation}
Consequently, the total disruption (ionization) probability during the time interval $(t,t+dt)$ is,
\begin{equation}
P_{\rm ion, \, i} = 1 - \exp\Big[-dt \big( \frac{1}{t_{\rm evap, \, i}} + \frac{1}{t_{\rm ejc, \, i}} \big)\Big].
\end{equation}
For each binary $j$, in each spherical shell $i$, and in each timestep $dt$, we draw a uniform random number $u_{i,j} \in [0,1]$. If $u_{i,j} < P_{\textrm{ion, $i$},j}$, binary $j$ is ionized during that timestep. The number of ionizations in shell $i$ over the interval  $(t, t+dt)$ in the sample is,
\begin{equation}
N_{\textrm{ion}, \, i}(t) = \sum_j \textrm{ionized}_{i,j},
\end{equation}
where $\textrm{ionized}_{i,j} = 1$ if $u_{i,j} < P_{\textrm{ion, $i$},j}$, and zero otherwise. The probability of a soft binary's survival at any given time and shell is $1-P_{\textrm{ion}, \, i}$.

\section{Halo effects on populations of merging PBH binaries}\label{halo_effect}

To study the effects of halos on the population of PBH binaries and their merger time, we select a sample of $N_{\rm surv}(z=12) = 4.75 \times 10^{6}$ PBH binaries that formed at $z = 3400$ and have survived GW mergers up to $z = 12$. We  first evolve this sample assuming the binaries remain unperturbed and evolve via GWs only using Eq.\eqref{eq:Evol_a_gw} to redshift $ z=0$, corresponding to the evolution time of $t_{\rm avaible}(z=12)=13.43$Gyr. For each  binary that merges within this time range, we record its merging time $t_{\rm merger}$ and the initial orbital parameters $(a_{\rm {init}}, e_{\rm init})$.
The result is a population of binaries that merge through pure GW evolution is shown in the top panel of Fig.~\ref{fig:merger_scatter1}, which presents a scatter plot of the initial semi-major axis $a_{\rm init}$ versus $1-e_{\rm init}$ for all binaries that merge by today, along with the corresponding merger time distribution. Note that for the top panel of Fig.~\ref{fig:merger_scatter1}, we assume $f_{\rm outside}(t)=1$ throughout all time. 

To study the effects of halos, we take this same surviving sample  $N_{\rm surv}(z=12) = 4.75 \times 10^{6}$  and identify the binaries that merge within 100 times the age of the universe if they evolve in isolation. 
We identify the binaries that merge within this time window to get a large population of binaries ($2.49 \times 10^{5}$) to evolve in the different environments out of the total $N_{\rm surv}(z=12)$. 
Our simulations study separately the evolution of the $2.49 \times 10^{5}$ binaries from redshift 12 to 0, on each spherical shell.
To keep track of the proper fraction of binaries that may merge, we take the entire population of $4.75 \times 10^{6}$ binaries and calculate when they enter a given shell in a given halo.
We compute the probability of a binary (from the $4.75 \times 10^{6}$ binaries) to enter a simulated shell $i$ of a halo of mass $M$, as a function of time. 

We define the probability of a PBH binary entering a shell $i$ with halo mass $M_{i}$ ($M= \sum_{i} M_{i}$), during $dt+t$ as, 
\begin{eqnarray}
\label{P_enter}
p_{\rm enter}(t+dt, M_{i}) &=& 
\frac{\Delta f_{\rm inside}(M_{i}, \, t+dt)}{1 - f_{\rm  inside}(M_{i}, \, t)} \\
&=& \frac{f_{\rm inside}(M_{i}, \, t+dt) - f_{\rm inside}(M_{i}, \, t)}
{1 - f_{\rm  inside}(M_{i}, \, t)}. \nonumber
\end{eqnarray}
The $f_{\rm inside}(M_{i}, \, t)$ is the probability of PBH binaries  being inside a shell of mass $M_{i}$ at time $t$,
\begin{equation}
f_{\rm inside}(M_{i}, \, t)= \frac{M_{i}(t[z])}{M_{i}(t[0])}.
\end{equation} 
Note that the normalization by $1- f_{\rm inside}(t)$ on the probability operator $p_{\rm enter}$, makes sure that only binaries currently outside halos are considered to enter 
\footnote{The number of binaries that by the end of the simulation are going to be inside a halo mass shell $i$ is less than $4.75 \times 10^{6}$ as mergers will happen before they enter, and at any earlier time they are proportional to ($1- f_{\rm inside}(t))$. As we describe in the main text, once evaluating the merger rate per halo we properly re-weight the merger events to account for the real amount of dark matter mass in each halo shell (see Eqs.~\eqref{eq:SiE}-\eqref{Rate_per_halo}).}. For each binary in the population at a given time, we draw a uniform random number $u_{i} \in (0,1)$) and use $p_{\rm enter}(t+dt)$ to decide whether it enters a halo, i.e if $u_{i}<p_{\rm enter}$, the binary $i$ in the population enters the halo, else , it stays outside.

When Binaries are outside halos, they evolve via GW emission only (Eq.~\eqref{eq:Evol_a_gw}), but when they enter inside a halo, they evolve according to Eqs.~\eqref{eq:evol_a_hard}--\eqref{eq:evol_e_soft}, depending on their type with respect to the halo and when they enter the halo as per Eq.~\eqref{P_enter}. 
\begin{figure}
    \centering
    \includegraphics[width=0.99\linewidth]{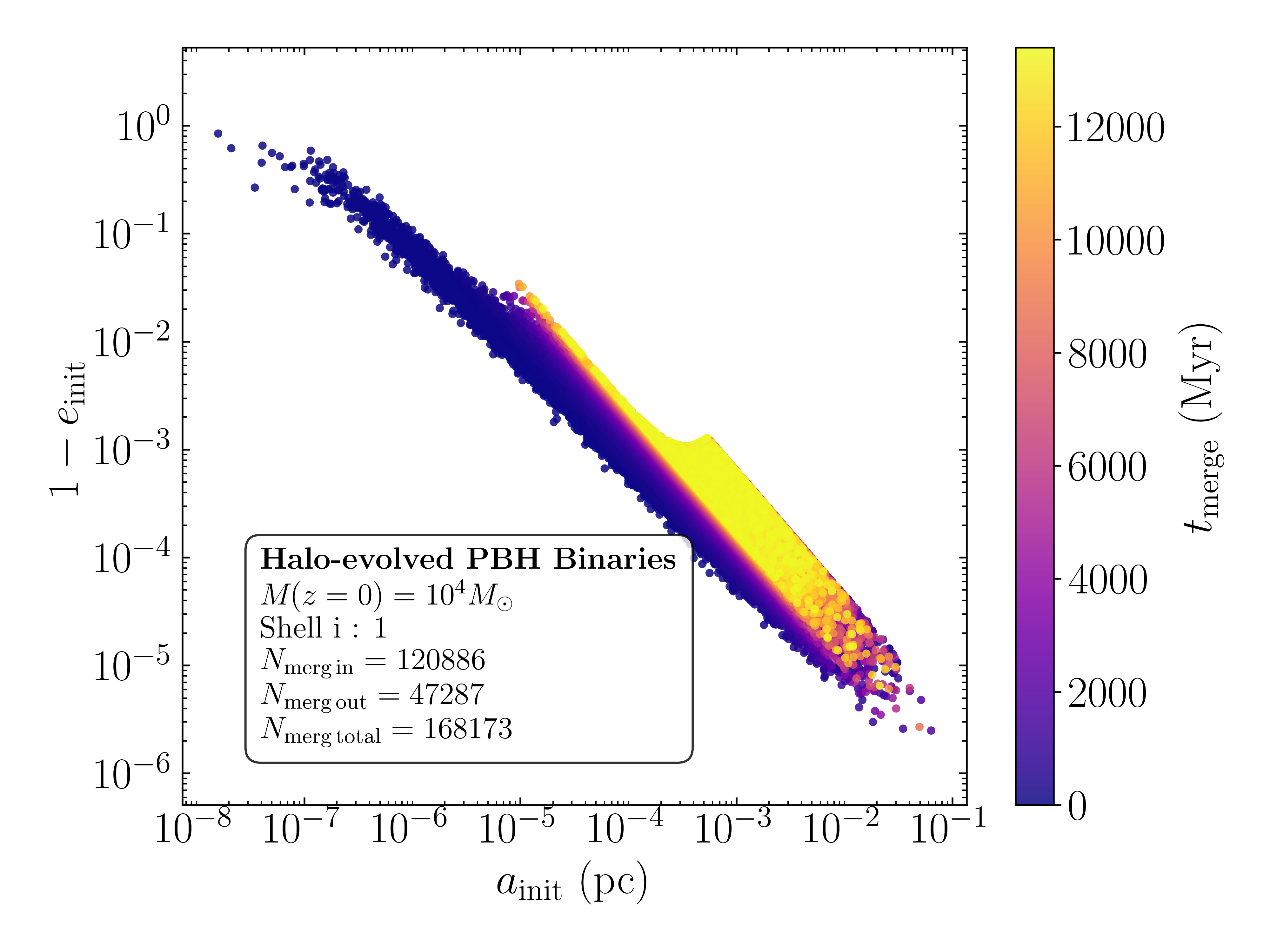}
    \vspace{-0.3cm}
    \includegraphics[width=0.99\linewidth]{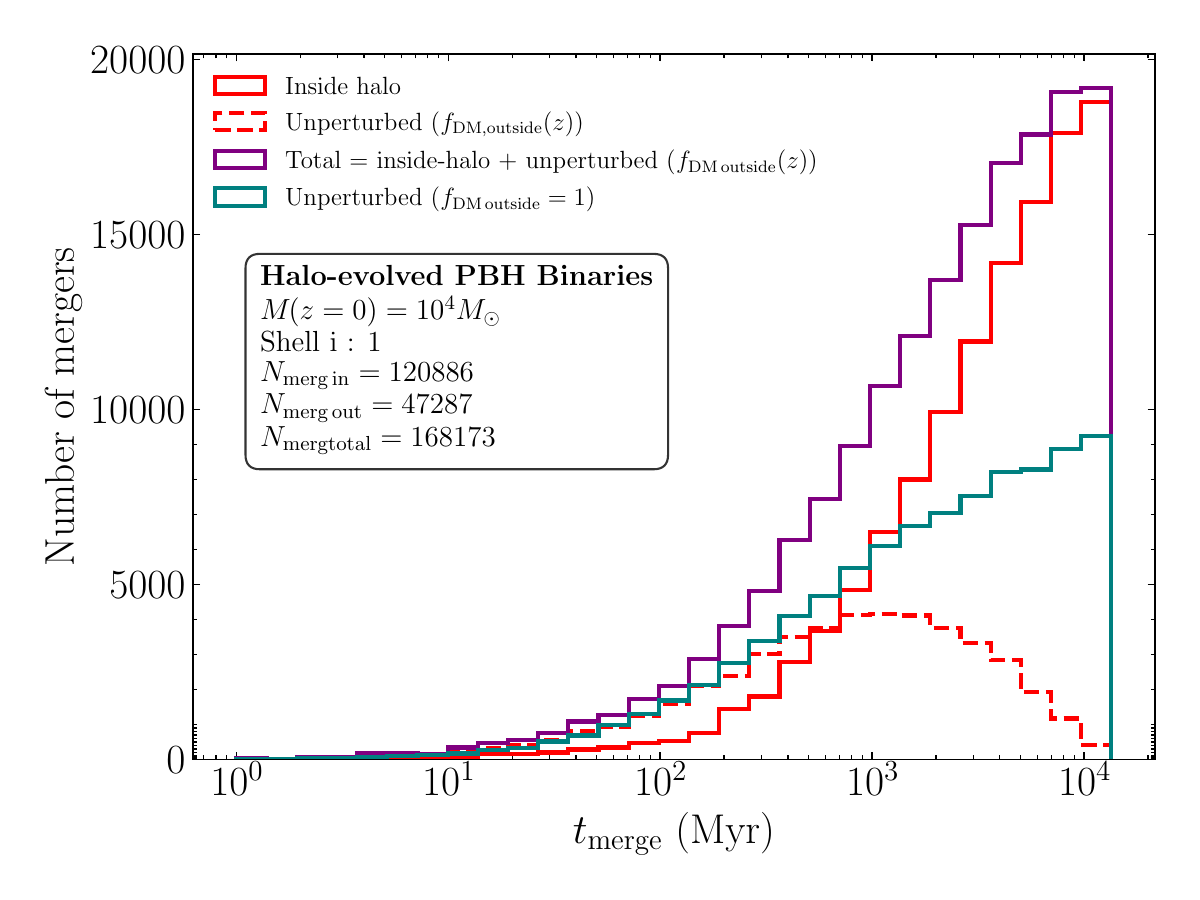}
    \vspace{-0.3cm}
    \includegraphics[width=0.99\linewidth]{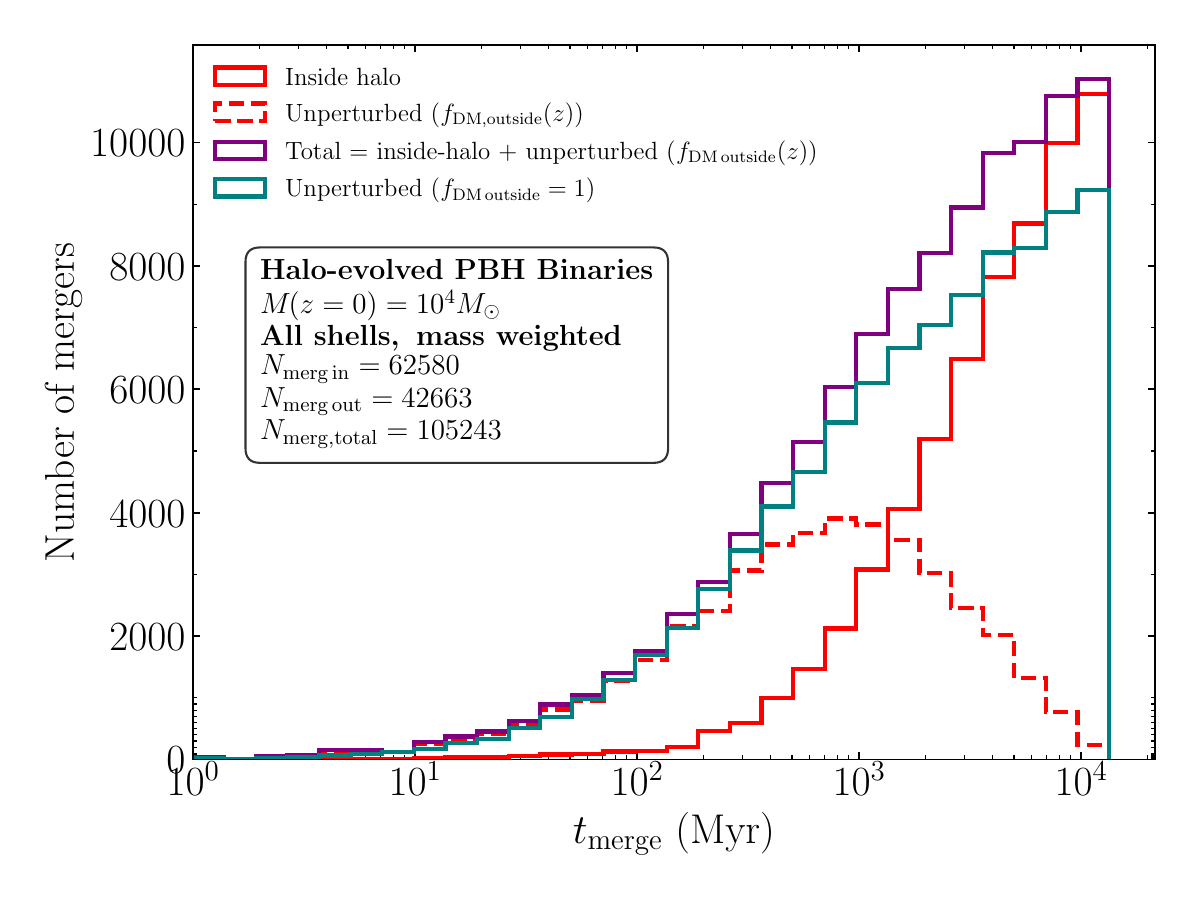}
    \vspace{-0.4cm}
    \caption{Same as in Figs.~\ref{fig:merger_scatter1} and~\ref{fig:merger_hist}, but for a $10^{4} \, M_{\odot}$ halo environment. In such halos, the merger rates of PBH binaries is enhanced.}
    \label{fig:Scatter_and_histrogram_10to4_Halo}
\end{figure}

\begin{figure}
    \centering
    \includegraphics[width=0.99\linewidth]{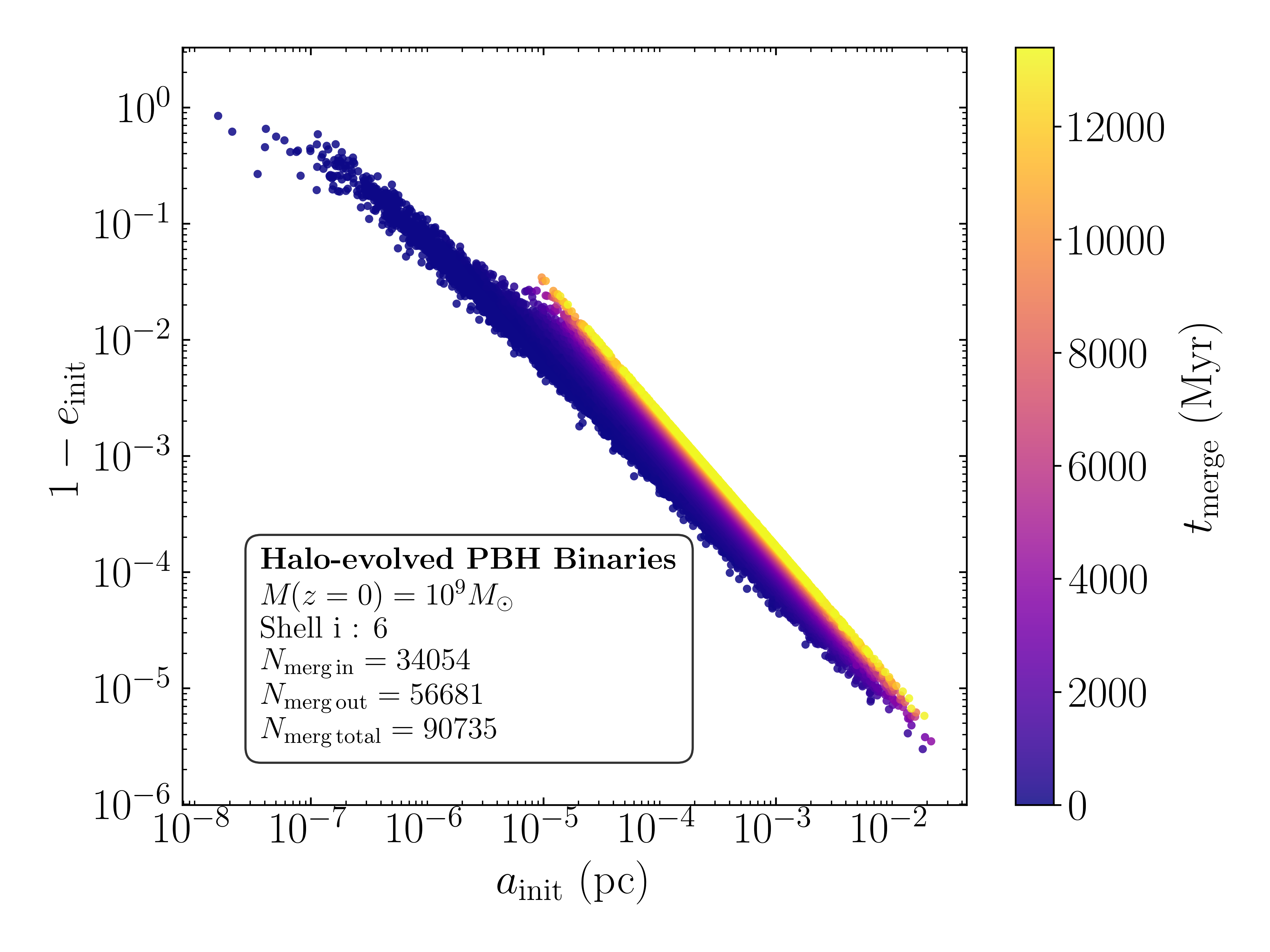}
    \vspace{-0.3cm}
    \includegraphics[width=0.99\linewidth]{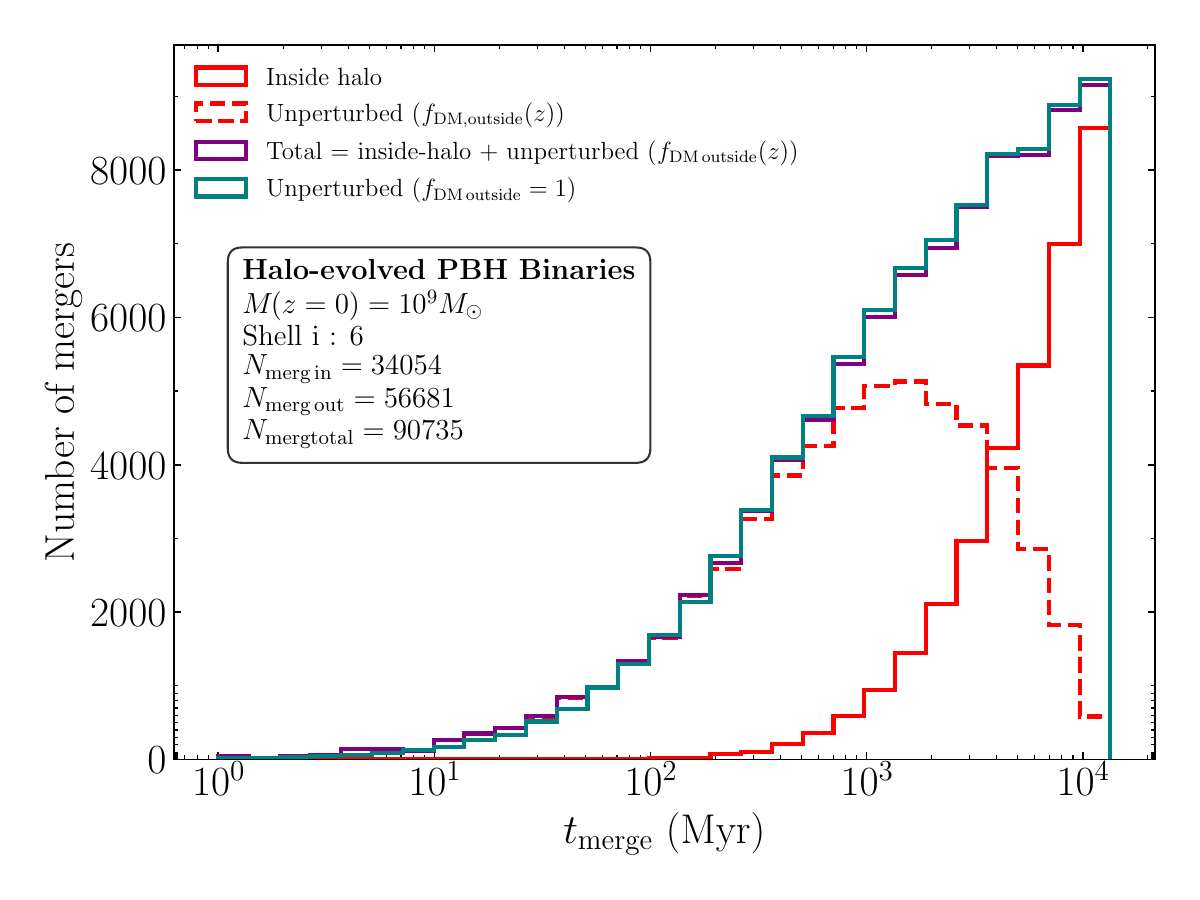}
    \vspace{-0.3cm}
    \includegraphics[width=0.99\linewidth]{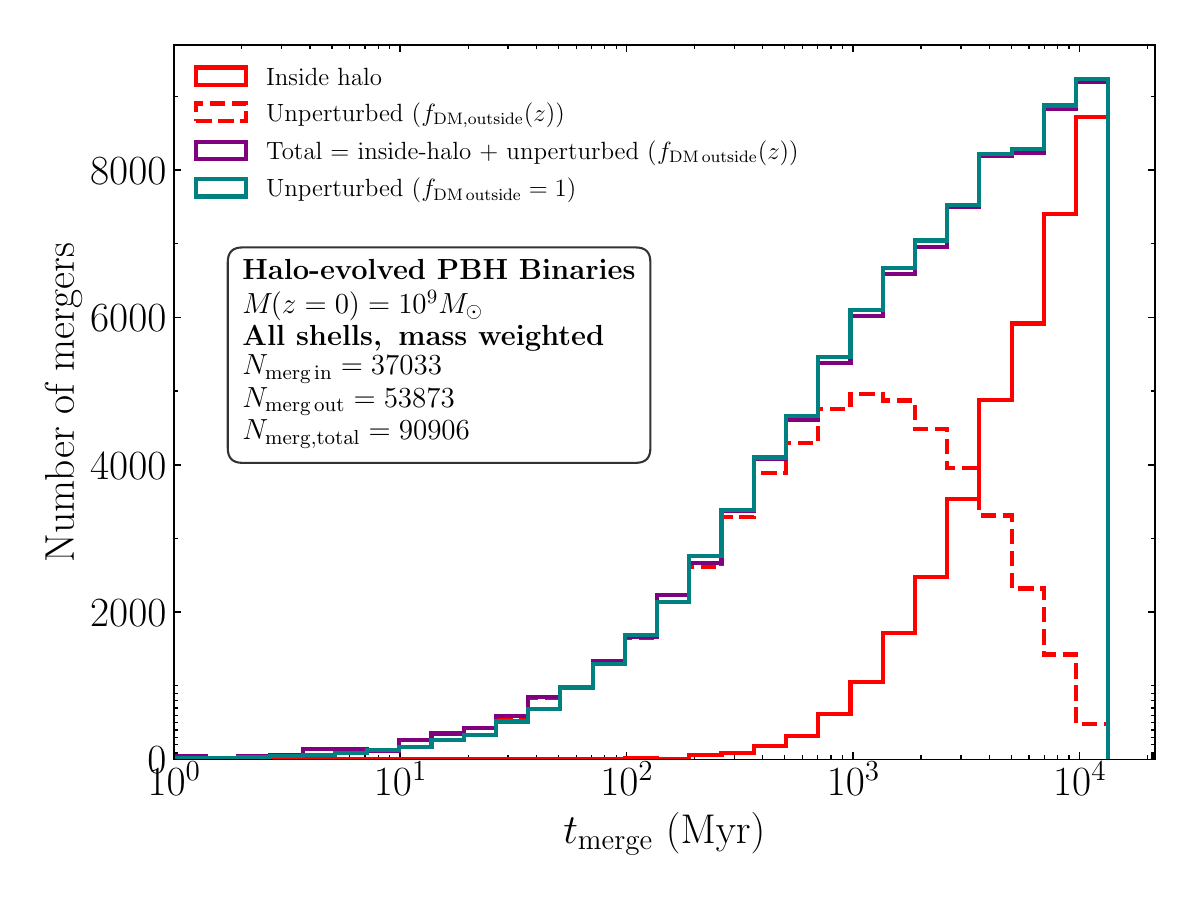}
    \vspace{-0.4cm}
    \caption{Same as in Figs.~\ref{fig:merger_scatter1} and~\ref{fig:merger_hist}, but for a $10^{9} \, M_{\odot}$ halo environment. In such halos, the merger rates of PBH binaries is essentially the same as that of unperturbed binaries.} \label{fig:Scatter_and_histrogram_10to9_Halo}
\end{figure}

In the top panels of Figs.~\ref{fig:Scatter_and_histrogram_10to4_Halo} and~\ref{fig:Scatter_and_histrogram_10to9_Halo}, we show the same scatter distribution of initial binary population properties as Fig.~\ref{fig:merger_scatter1} of the main text, but now assuming instead the halo environmental conditions of a host dark matter halo of $M(z=0) = 10^{4}  \, M_{\odot}$ and a $M(z=0) =  10^{9} \, M_{\odot}$ halo respectively. In the middle panels of Figs.~\ref{fig:Scatter_and_histrogram_10to4_Halo} and~\ref{fig:Scatter_and_histrogram_10to9_Halo}, we give the  equivalent histograms to the top panel of Fig.~\ref{fig:merger_hist}, but for the $10^{4}$ and the $10^{9}$ $M_{\odot}$ dark matter halos respectively.
For the middle panels of Figs.~\ref{fig:Scatter_and_histrogram_10to4_Halo} and~\ref{fig:Scatter_and_histrogram_10to9_Halo}, we take the same environmental assumptions used to get the scatter plots of their respective top panels.
Finally, in the bottom panels of Figs.~\ref{fig:Scatter_and_histrogram_10to4_Halo} and~\ref{fig:Scatter_and_histrogram_10to9_Halo}, we give the shell mass-weighted merger events for the $10^{4}$ and the $10^{9}$ $M_{\odot}$ dark matter halos. 
Binary-single interactions are more important in smaller mass halos as the dynamical terms of Eqs.~\eqref{eq:evol_a_hard} and~\eqref{eq:evol_e_soft} do affect the merger timescale. Instead, in the most massive dark matter halos, the unperturbed merger rate assumption is effectively correct. Due to the stochastic nature of selecting the orbital properties of the PBH binaries at their formation from their underlying distribution, the exact number of merged binaries between $z=12$ and $z=0$, may change minimally between simulations. In Fig.~\ref{fig:merger_scatter1}, and in the green histograms throughout, there are 90830 binaries that merge in that interval, in isolation, i.e. unperturbed from third objects. In the panels of Fig.~\ref{fig:Scatter_and_histrogram_10to9_Halo}, the number of mergers (90735 and 90906) inside the halos is essentially the same as that of unperturbed binaries. 

\section{The PBH binary population and its sampling procedure} \label{sample_update}
At each timestep $(t,t+dt)$ of our simulations the actual number of PBH binaries inside halo shell $i$, $N_{\textrm{PBH binaries},i}(t)$, is updated as follows: (1) binaries that survive merging or getting ionized during the interval $(t, t+dt)$ are retained, and  (2)  newly added  binaries $N_{\textrm{new PBH binaries},i}(t+dt)$ outside halos enter shell $i$ due to growth in the shell (and overall halo) mass. This translates to,
\begin{equation}
\begin{split}
N_{\textrm{PBH binaries},i}(t+dt) = & S_i^{\textrm{surv}}(t)\, N_{\textrm{PBH binaries},i}(t) \\
& + N_{\textrm{new PBH binaries},i}(t+dt),
\end{split}
\end{equation}
where $S_i^{\textrm{surv}}(t)$ is the probability of survival of all binaries (hard and soft) from either mergers or ionizations. 
Meanwhile, to make sure the sample in shell $i$, $N_{\textrm{sample},i}(t)$, stays accurate and reflects the two populations of binaries in each timestep, the evolving sample is split into survived and newly added binaries according to their relative fractions in each shell $i$.
The fraction of new binaries in shell $i$ at time $t+dt$ is,  
\begin{equation}
f_{\textrm{new},i}(t+dt) = \frac{N_{\textrm{new PBH binaries},i}(t+dt)}{N_{\textrm{PBH binaries},i}(t+dt)}.
\end{equation}
%Thus, the splitting of the PBHs sampling is done according to,
%\begin{eqnarray}
%N_{\textrm{new sample},i} &=& %f_{\textrm{new},i}(t+dt) \times N_{\textrm{sample},i}, \\
%N_{\textrm{survived sample},i} &=& N_{\textrm{sample},i} - N_{\textrm{new sample},i}.
%\end{eqnarray}

The survived sample is drawn from binaries that did not merge or get ionized during the previous timestep in shell $i$. The new sample is drawn from unperturbed (i.e, isolated) binaries that in the previous timestep were outside the halo. This approach  makes sure that our sample accurately reflects the evolving binary population within each halo shell.

\bibliography{REF.bib}

@article{Hawking:1971ei,
    author = "Hawking, Stephen",
    title = "{Gravitationally collapsed objects of very low mass}",
    doi = "10.1093/mnras/152.1.75",
    journal = "Mon. Not. Roy. Astron. Soc.",
    volume = "152",
    pages = "75",
    year = "1971"
}

@article{Zeldovich:1967lct,
    author = "Zel'dovich, Ya. B. and Novikov, I. D.",
    title = "{The Hypothesis of Cores Retarded during Expansion and the Hot Cosmological Model}",
    journal = "Sov. Astron.",
    volume = "10",
    pages = "602",
    year = "1967"
}

@article{Planck:2018vyg,
    author = "Aghanim, N. and others",
    collaboration = "Planck",
    title = "{Planck 2018 results. VI. Cosmological parameters}",
    eprint = "1807.06209",
    archivePrefix = "arXiv",
    primaryClass = "astro-ph.CO",
    doi = "10.1051/0004-6361/201833910",
    journal = "Astron. Astrophys.",
    volume = "641",
    pages = "A6",
    year = "2020",
    note = "[Erratum: Astron.Astrophys. 652, C4 (2021)]"
}

@article{KAGRA:2021duu,
    author = "Abbott, R. and others",
    collaboration = "KAGRA, VIRGO, LIGO Scientific",
    title = "{Population of Merging Compact Binaries Inferred Using Gravitational Waves through GWTC-3}",
    eprint = "2111.03634",
    archivePrefix = "arXiv",
    primaryClass = "astro-ph.HE",
    reportNumber = "LIGO-P2100239 ; Data release: https://zenodo.org/record/5655785, LIGO-P2100239",
    doi = "10.1103/PhysRevX.13.011048",
    journal = "Phys. Rev. X",
    volume = "13",
    number = "1",
    pages = "011048",
    year = "2023"
}

@article{KAGRA:2021vkt,
    author = "Abbott, R. and others",
    collaboration = "KAGRA, VIRGO, LIGO Scientific",
    title = "{GWTC-3: Compact Binary Coalescences Observed by LIGO and Virgo during the Second Part of the Third Observing Run}",
    eprint = "2111.03606",
    archivePrefix = "arXiv",
    primaryClass = "gr-qc",
    reportNumber = "LIGO-P2000318",
    doi = "10.1103/PhysRevX.13.041039",
    journal = "Phys. Rev. X",
    volume = "13",
    number = "4",
    pages = "041039",
    year = "2023"
}

@article{Kovetz:2016kpi,
    author = "Kovetz, Ely D. and Cholis, Ilias and Breysse, Patrick C. and Kamionkowski, Marc",
    title = "{Black hole mass function from gravitational wave measurements}",
    eprint = "1611.01157",
    archivePrefix = "arXiv",
    primaryClass = "astro-ph.CO",
    doi = "10.1103/PhysRevD.95.103010",
    journal = "Phys. Rev. D",
    volume = "95",
    number = "10",
    pages = "103010",
    year = "2017"
}

@article{Takhistov:2017bpt,
    author = "Takhistov, Volodymyr",
    title = "{Transmuted Gravity Wave Signals from Primordial Black Holes}",
    eprint = "1707.05849",
    archivePrefix = "arXiv",
    primaryClass = "astro-ph.CO",
    doi = "10.1016/j.physletb.2018.05.026",
    journal = "Phys. Lett. B",
    volume = "782",
    pages = "77--82",
    year = "2018"
}

@article{Kritos:2020fjw,
    author = "Kritos, Konstantinos and Cholis, Ilias",
    title = "{Evaluating the merger rate of binary black holes from direct captures and third-body soft interactions using the Milky Way globular clusters}",
    eprint = "2007.02968",
    archivePrefix = "arXiv",
    primaryClass = "astro-ph.GA",
    doi = "10.1103/PhysRevD.102.083016",
    journal = "Phys. Rev. D",
    volume = "102",
    number = "8",
    pages = "083016",
    year = "2020"
}

@article{Kritos:2022ggc,
    author = "Kritos, Konstantinos and Strokov, Vladimir and Baibhav, Vishal and Berti, Emanuele",
    title = "{Dynamical formation of black hole binaries in dense star clusters: Rapid cluster evolution code}",
    eprint = "2210.10055",
    archivePrefix = "arXiv",
    primaryClass = "astro-ph.HE",
    doi = "10.1103/PhysRevD.110.043023",
    journal = "Phys. Rev. D",
    volume = "110",
    number = "4",
    pages = "043023",
    year = "2024"
}

@article{Sesana:2006xw,
    author = "Sesana, Alberto and Haardt, Francesco and Madau, Piero",
    title = "{Interaction of massive black hole binaries with their stellar environment. 1. Ejection of hypervelocity stars}",
    eprint = "astro-ph/0604299",
    archivePrefix = "arXiv",
    doi = "10.1086/507596",
    journal = "Astrophys. J.",
    volume = "651",
    pages = "392--400",
    year = "2006"
}

@ARTICLE{1964PhRv..136.1224P,
       author = {{Peters}, P.~C.},
        title = "{Gravitational Radiation and the Motion of Two Point Masses}",
      journal = {Physical Review},
         year = 1964,
        month = nov,
       volume = {136},
       number = {4B},
        pages = {1224-1232},
          doi = {10.1103/PhysRev.136.B1224},
       adsurl = {https://ui.adsabs.harvard.edu/abs/1964PhRv..136.1224P}
}

@article{Kovetz:2017rvv,
    author = "Kovetz, Ely D.",
    title = "{Probing Primordial-Black-Hole Dark Matter with Gravitational Waves}",
    eprint = "1705.09182",
    archivePrefix = "arXiv",
    primaryClass = "astro-ph.CO",
    doi = "10.1103/PhysRevLett.119.131301",
    journal = "Phys. Rev. Lett.",
    volume = "119",
    number = "13",
    pages = "131301",
    year = "2017"
}

@article{Bellomo:2017zsr,
    author = "Bellomo, Nicola and Bernal, Jos{\'e} Luis and Raccanelli, Alvise and Verde, Licia",
    title = "{Primordial Black Holes as Dark Matter: Converting Constraints from Monochromatic to Extended Mass Distributions}",
    eprint = "1709.07467",
    archivePrefix = "arXiv",
    primaryClass = "astro-ph.CO",
    doi = "10.1088/1475-7516/2018/01/004",
    journal = "JCAP",
    volume = "01",
    pages = "004",
    year = "2018"
}

@article{Raccanelli:2016cud,
    author = "Raccanelli, Alvise and Kovetz, Ely D. and Bird, Simeon and Cholis, Ilias and Munoz, Julian B.",
    title = "{Determining the progenitors of merging black-hole binaries}",
    eprint = "1605.01405",
    archivePrefix = "arXiv",
    primaryClass = "astro-ph.CO",
    doi = "10.1103/PhysRevD.94.023516",
    journal = "Phys. Rev. D",
    volume = "94",
    number = "2",
    pages = "023516",
    year = "2016"
}

@article{Bouhaddouti:2024ena,
    author = "Bouhaddouti, Mehdi El and Cholis, Ilias",
    title = "{On the mass distribution of the LIGO-Virgo-KAGRA events}",
    eprint = "2409.00179",
    archivePrefix = "arXiv",
    primaryClass = "astro-ph.CO",
    doi = "10.1103/PhysRevD.111.043020",
    journal = "Phys. Rev. D",
    volume = "111",
    number = "4",
    pages = "043020",
    year = "2025"
}

@article{Berti:2019xgr,
    author = "Berti, Emanuele and others",
    title = "{Tests of General Relativity and Fundamental Physics with Space-based Gravitational Wave Detectors}",
    eprint = "1903.02781",
    archivePrefix = "arXiv",
    primaryClass = "astro-ph.HE",
    month = "3",
    year = "2019"
}

@article{Gow:2019pok,
    author = "Gow, Andrew D. and Byrnes, Christian T. and Hall, Alex and Peacock, John A.",
    title = "{Primordial black hole merger rates: distributions for multiple LIGO observables}",
    eprint = "1911.12685",
    archivePrefix = "arXiv",
    primaryClass = "astro-ph.CO",
    doi = "10.1088/1475-7516/2020/01/031",
    journal = "JCAP",
    volume = "01",
    pages = "031",
    year = "2020"
}

@article{Andres-Carcasona:2024wqk,
    author = {Andr{\'e}s-Carcasona, M. and Iovino, A. J. and Vaskonen, V. and Veerm{\"a}e, H. and Mart{\'\i}nez, M. and Pujol{\`a}s, O. and Mir, Ll. M.},
    title = "{Constraints on primordial black holes from LIGO-Virgo-KAGRA O3 events}",
    eprint = "2405.05732",
    archivePrefix = "arXiv",
    primaryClass = "astro-ph.CO",
    doi = "10.1103/PhysRevD.110.023040",
    journal = "Phys. Rev. D",
    volume = "110",
    number = "2",
    pages = "023040",
    year = "2024"
}

@article{Hall:2020daa,
    author = "Hall, Alex and Gow, Andrew D. and Byrnes, Christian T.",
    title = "{Bayesian analysis of LIGO-Virgo mergers: Primordial vs. astrophysical black hole populations}",
    eprint = "2008.13704",
    archivePrefix = "arXiv",
    primaryClass = "astro-ph.CO",
    doi = "10.1103/PhysRevD.102.123524",
    journal = "Phys. Rev. D",
    volume = "102",
    pages = "123524",
    year = "2020"
}

@article{Bouhaddouti:2025ltb,
    author = "Bouhaddouti, Mehdi El and Aljaf, Muhsin and Cholis, Ilias",
    title = "{Conservative limits on primordial black holes from the LIGO-Virgo-KAGRA observations}",
    eprint = "2502.00144",
    archivePrefix = "arXiv",
    primaryClass = "astro-ph.CO",
    month = "1",
    year = "2025"
}

@article{Carr:2016drx,
    author = "Carr, Bernard and Kuhnel, Florian and Sandstad, Marit",
    title = "{Primordial Black Holes as Dark Matter}",
    eprint = "1607.06077",
    archivePrefix = "arXiv",
    primaryClass = "astro-ph.CO",
    reportNumber = "NORDITA-2016-83",
    doi = "10.1103/PhysRevD.94.083504",
    journal = "Phys. Rev. D",
    volume = "94",
    number = "8",
    pages = "083504",
    year = "2016"
}

@article{LIGOScientific:2016aoc,
    author = "Abbott, B. P. and others",
    collaboration = "LIGO Scientific, Virgo",
    title = "{Observation of Gravitational Waves from a Binary Black Hole Merger}",
    eprint = "1602.03837",
    archivePrefix = "arXiv",
    primaryClass = "gr-qc",
    reportNumber = "LIGO-P150914",
    doi = "10.1103/PhysRevLett.116.061102",
    journal = "Phys. Rev. Lett.",
    volume = "116",
    number = "6",
    pages = "061102",
    year = "2016"
}

@article{Sasaki:2016jop,
    author = "Sasaki, Misao and Suyama, Teruaki and Tanaka, Takahiro and Yokoyama, Shuichiro",
    title = "{Primordial Black Hole Scenario for the Gravitational-Wave Event GW150914}",
    eprint = "1603.08338",
    archivePrefix = "arXiv",
    primaryClass = "astro-ph.CO",
    reportNumber = "RESCEU-17-16, RUP-16-7, YITP-16-43",
    doi = "10.1103/PhysRevLett.117.061101",
    journal = "Phys. Rev. Lett.",
    volume = "117",
    number = "6",
    pages = "061101",
    year = "2016",
    note = "[Erratum: Phys.Rev.Lett. 121, 059901 (2018)]"
}

@article{Bird:2016dcv,
    author = {Bird, Simeon and Cholis, Ilias and Mu{\~n}oz, Julian B. and Ali-Ha{\"\i}moud, Yacine and Kamionkowski, Marc and Kovetz, Ely D. and Raccanelli, Alvise and Riess, Adam G.},
    title = "{Did LIGO detect dark matter?}",
    eprint = "1603.00464",
    archivePrefix = "arXiv",
    primaryClass = "astro-ph.CO",
    doi = "10.1103/PhysRevLett.116.201301",
    journal = "Phys. Rev. Lett.",
    volume = "116",
    number = "20",
    pages = "201301",
    year = "2016"
}

@article{Ali-Haimoud:2017rtz,
    author = {Ali-Ha{\"\i}moud, Yacine and Kovetz, Ely D. and Kamionkowski, Marc},
    title = "{Merger rate of primordial black-hole binaries}",
    eprint = "1709.06576",
    archivePrefix = "arXiv",
    primaryClass = "astro-ph.CO",
    doi = "10.1103/PhysRevD.96.123523",
    journal = "Phys. Rev. D",
    volume = "96",
    number = "12",
    pages = "123523",
    year = "2017"
}

@article{Franciolini:2021xbq,
    author = "Franciolini, Gabriele and Cotesta, Roberto and Loutrel, Nicholas and Berti, Emanuele and Pani, Paolo and Riotto, Antonio",
    title = "{How to assess the primordial origin of single gravitational-wave events with mass, spin, eccentricity, and deformability measurements}",
    eprint = "2112.10660",
    archivePrefix = "arXiv",
    primaryClass = "astro-ph.CO",
    reportNumber = "ET-0464A-21",
    doi = "10.1103/PhysRevD.105.063510",
    journal = "Phys. Rev. D",
    volume = "105",
    number = "6",
    pages = "063510",
    year = "2022"
}

@article{Kavanagh:2018ggo,
    author = "Kavanagh, Bradley J. and Gaggero, Daniele and Bertone, Gianfranco",
    title = "{Merger rate of a subdominant population of primordial black holes}",
    eprint = "1805.09034",
    archivePrefix = "arXiv",
    primaryClass = "astro-ph.CO",
    doi = "10.1103/PhysRevD.98.023536",
    journal = "Phys. Rev. D",
    volume = "98",
    number = "2",
    pages = "023536",
    year = "2018"
}

@inbook{Raidal:2024bmm,
    author = {Raidal, Martti and Vaskonen, Ville and Veerm{\"a}e, Hardi},
    editor = "Byrnes, Christian and Franciolini, Gabriele and Harada, Tomohiro and Pani, Paolo and Sasaki, Misao",
    title = "{Formation of~Primordial Black Hole Binaries and~Their Merger Rates}",
    booktitle = "{Primordial Black Holes}",
    eprint = "2404.08416",
    archivePrefix = "arXiv",
    primaryClass = "astro-ph.CO",
    doi = "10.1007/978-981-97-8887-3_16",
    year = "2025"
}

@article{Cholis:2016kqi,
    author = {Cholis, Ilias and Kovetz, Ely D. and Ali-Ha{\"\i}moud, Yacine and Bird, Simeon and Kamionkowski, Marc and Mu{\~n}oz, Julian B. and Raccanelli, Alvise},
    title = "{Orbital eccentricities in primordial black hole binaries}",
    eprint = "1606.07437",
    archivePrefix = "arXiv",
    primaryClass = "astro-ph.HE",
    doi = "10.1103/PhysRevD.94.084013",
    journal = "Phys. Rev. D",
    volume = "94",
    number = "8",
    pages = "084013",
    year = "2016"
}

@article{Press:1973iz,
    author = "Press, William H. and Schechter, Paul",
    title = "{Formation of galaxies and clusters of galaxies by selfsimilar gravitational condensation}",
    doi = "10.1086/152650",
    journal = "Astrophys. J.",
    volume = "187",
    pages = "425--438",
    year = "1974"
}

@article{Aljaf:2024fru,
  title = {Simulating binary primordial black hole mergers in dark matter halos},
  author = {Aljaf, Muhsin and Cholis, Ilias},
  journal = {Phys. Rev. D},
  volume = {111},
  issue = {6},
  pages = {063020},
  numpages = {19},
  year = {2025},
  month = {Mar},
  publisher = {American Physical Society},
  doi = {10.1103/PhysRevD.111.063020},
  url = {https://link.aps.org/doi/10.1103/PhysRevD.111.063020}
}

@article{Correa:2015kia,
    author = "Correa, Camila A. and Wyithe, J. Stuart B. and Schaye, Joop and Duffy, Alan R.",
    title = "{The accretion history of dark matter haloes {\textendash} II. The connections with the mass power spectrum and the density profile}",
    eprint = "1501.04382",
    archivePrefix = "arXiv",
    primaryClass = "astro-ph.CO",
    doi = "10.1093/mnras/stv697",
    journal = "Mon. Not. Roy. Astron. Soc.",
    volume = "450",
    number = "2",
    pages = "1521--1537",
    year = "2015"
}

@article{Heggie:1975tg,
    author = "Heggie, D. C.",
    title = "{Binary evolution in stellar dynamics}",
    journal = "Mon. Not. Roy. Astron. Soc.",
    volume = "173",
    pages = "729--787",
    year = "1975"
}

@article{Navarro:1995iw,
    author = "Navarro, Julio F. and Frenk, Carlos S. and White, Simon D. M.",
    title = "{The Structure of cold dark matter halos}",
    eprint = "astro-ph/9508025",
    archivePrefix = "arXiv",
    doi = "10.1086/177173",
    journal = "Astrophys. J.",
    volume = "462",
    pages = "563--575",
    year = "1996"
}

@article{Murray:2013qza,
    author = "Murray, Steven and Power, Chris and Robotham, A. S. G.",
    title = "{HMFcalc: An online tool for calculating dark matter halo mass functions}",
    eprint = "1306.6721",
    archivePrefix = "arXiv",
    primaryClass = "astro-ph.CO",
    doi = "10.1016/j.ascom.2013.11.001",
    journal = "Astron. Comput.",
    volume = "3-4",
    pages = "23--34",
    year = "2013"
}

@article{Quinlan:1996vp,
    author = "Quinlan, Gerald D.",
    title = "{The dynamical evolution of massive black hole binaries - I. hardening in a fixed stellar background}",
    eprint = "astro-ph/9601092",
    archivePrefix = "arXiv",
    reportNumber = "RUTGERS-ASTROPHYSICS-PREPRINT-SERIES-NO-187",
    doi = "10.1016/S1384-1076(96)00003-6",
    journal = "New Astron.",
    volume = "1",
    pages = "35--56",
    year = "1996"
}

@BOOK{1987Binney,
       author = {{Binney}, James and {Tremaine}, Scott},
        title = "{Galactic dynamics}",
         year = 1987,
       adsurl = {https://ui.adsabs.harvard.edu/abs/1987gady.book.....B}
}
\end{document}